\documentclass[rmp, aps, floatfix, nofootinbib, twocolumn]{revtex4-1}

\usepackage{amsmath,amsfonts,amssymb,bm}
\usepackage{dsfont}
\usepackage{graphicx}
\usepackage{color}
\usepackage{soul}
\usepackage{slashed}

\def\inbar{\,\vrule height1.5ex width.4pt depth0pt}
\def\IR{\relax{\rm I\kern-.18em R}}
\def\IC{\relax\hbox{$\inbar\kern-.3em{\rm C}$}}

\usepackage[mathscr,scaled=1.15]{urwchancal}
\DeclareFontFamily{OT1}{pzc}{}
\DeclareFontShape{OT1}{pzc}{m}{it}%
{<-> s * [1.15] pzcmi7t}{}
\DeclareMathAlphabet{\mathpzc}{OT1}{pzc}{m}{it}

\definecolor{purple}{rgb}{0.5,0,0.5}
\definecolor{blue}{rgb}{0.0,0,0.9}
\definecolor{prdblue}{rgb}{0.133,0.118,0.498}
\usepackage[colorlinks=true, pdfstartview=FitV, linkcolor=prdblue, citecolor= prdblue, urlcolor=prdblue]{hyperref}

\begin{document}
\title{Roper resonance -- solution to the fifty year puzzle}

\author{Volker D. Burkert}
\email{burkert@jlab.org}
\affiliation{Thomas Jefferson National Accelerator Facility, Newport News, Virginia 23606, USA}

\author{Craig D. Roberts}
\email{cdroberts@anl.gov}
\affiliation{Physics Division, Argonne National Laboratory, Argonne, Illinois 60439, USA}

\renewcommand{\baselinestretch}{1.1}

\begin{abstract}
For half a century, the Roper resonance has defied understanding.  Discovered in 1963, it appears to be an exact copy of the proton \emph{except} that its mass is 50\% greater.
The mass is the first problem: it is difficult to explain with any theoretical tool that can validly be used to study the strong-interaction piece of the Standard Model of Particle Physics, \emph{i.e}.\ quantum chromodynamics [QCD].
%
In the last decade, a new challenge has appeared, \emph{viz}.\ precise information on the proton-to-Roper electroproduction transition form factors, reaching out to momentum transfer $Q^2 \approx 4.5\,$GeV$^2$.  This scale probes the domain within which hard valence-quark degrees-of-freedom could be expected to determine form factor behavior.  Hence, with this new data the Roper resonance becomes a problem for strong-QCD [sQCD].
An explanation of how and where the Roper resonance fits into the emerging spectrum of hadrons cannot rest on a description of its mass alone.  Instead, it must combine an understanding of the Roper's mass and width with a detailed account of its structure and how that structure is revealed in the momentum dependence of the transition form factors.  Furthermore, it must unify all this with a similarly complete picture of the proton from which the Roper resonance is produced.
This is a prodigious task, but a ten-year international collaborative effort, drawing together experimentalists and theorists, has presented a solution to the puzzle.  Namely, the observed Roper is at heart the proton's first radial excitation, consisting of a dressed-quark core augmented by a meson cloud that reduces the core mass by approximately 20\% and materially alters its electroproduction form factors on $Q^2 < 2\,m_N^2$, where $m_N$ is the proton's mass.
We describe the experimental motivations and developments which enabled electroproduction data to be procured within a domain that is unambiguously the purview of sQCD, thereby providing a real challenge and opportunity for modern theory; and survey the developments in reaction models and QCD theory that have enabled this conclusion to be drawn about the nature of the Roper resonance.
%
\end{abstract}
\date{21 September 2017}

\maketitle

\tableofcontents

\section{Introduction}
The Roper resonance was discovered in 1963 \cite{Roper:1964zza, BAREYRE1964137, AUVIL196476, PhysRevLett.13.555, PhysRev.138.B190}; and, as we shall relate, its characteristics have been the source of great puzzlement since that time.  It is therefore appropriate here to state the simplest of these characteristics; namely, the Roper is a $J=1/2$ positive-parity resonance with pole mass $\approx 1.37\,$GeV and width $\approx 0.18\,$GeV \cite{Olive:2016xmw}.  In the spectrum of nucleon-like states, \emph{i.e}.\ baryons with isospin\footnote{Isospin is a quantum number associated with strong-interaction bound-states.  Its value indicates the number of electric-charge states that may be considered as (nearly) identical in the absence of electroweak interactions, \emph{e.g}.\ the neutron and proton form an $I=1/2$ multiplet and are collectively described as nucleons.}  
$I=1/2$, the Roper resonance lies about $0.4\,$GeV above the ground-state nucleon and $0.15\,$GeV below the first $J=1/2$ negative-parity state, which has roughly the same width.  Today, the levels in this spectrum are labelled thus:
$N({\rm mass})\,J^P$;
and hence the ground-state nucleon is denoted $N(940)\,1/2^+$, the Roper resonance as $N(1440)\,1/2^+$, and the negative-parity state described above is $N(1535)\,1/2^-$.

The search for an understanding of the Roper resonance is the highest profile case in a long-running effort to chart and explain the spectrum and interactions of all strong interaction bound states that are supported by the Standard Model of Particle Physics.  The importance of this effort has long been recognized and cannot be overestimated.  Indeed, baryons and their resonances play a central role in the existence of our universe and ourselves; and therefore \cite{Isgur:2000ad}: ``\emph{\ldots\ they must be at the center of any discussion of why the world we actually experience has the character it does.  I am convinced that completing this chapter in the history of science will be one of the most interesting and fruitful areas of physics for at least the next thirty years.}''

Strong interactions within the Standard Model are described by quantum chromodynamics (QCD), the theory of gluons (gauge fields) and quarks (matter fields).   QCD is conceptually simple and can be expressed compactly in just one line, with two definitions \cite{Wilczek:2000ih}; and yet, nearly four decades after its formulation, we are still seeking answers to such apparently simple questions as what is the proton's wave function and which, if any, of the known baryons is the proton's first radial excitation.  Indeed, numerous problems remain open because QCD is fundamentally different from the Standard Model's other pieces: whilst a perturbation theory exists and is a powerful tool when used in connection with high-energy QCD processes, it is essentially useless when it comes to developing an understanding of strong interaction bound states built from light quarks.\footnote{There are six known quark flavors: $u$ (up) and $d$ (down) quarks are light, with masses far less than the characteristic QCD mass-scale of $\Lambda_{\rm QCD} \approx 0.2\,$GeV; $s$ (strange) quarks lie near the boundary between light and heavy; the $c$ (charm) quarks are relatively heavy, but not heavy enough for non-relativistic approximations to be quantitatively accurate; $b$ (bottom) quarks are practically heavy; and $t$ (top) quark are so heavy that they decay via weak interactions before forming hadron bound-states.}

The study of the properties and interactions of light hadronic systems lies squarely within the purview of strong-QCD [sQCD], \emph{viz}.\ the body of experimental and theoretical methods used to probe and map the infrared domain of Standard Model physics, whereupon emergent phenomena, such as gluon and quark confinement and dynamical chiral symmetry breaking [DCSB], appear to play the dominant role in determining all observable characteristics of the theory.  The nature of sQCD, and its contemporary methods and challenges will become apparent as we recount the history of the Roper resonance and the modern developments that have enabled a coherent picture of this system to emerge and, by analogy, of an array of related resonances.





\section{Constituent Quark Model Expectations}
Theoretical speculations on the nature of the Roper resonance followed immediately upon its discovery.  For instance, it was emphasized that the enhancement observed in experiment need not necessarily be identified with a resonant state \cite{DALITZ1965159}; but if it is a resonance, then it has structural similarities with the ground-state nucleon \cite{PhysRevLett.16.772}.

\begin{figure}[t]
\centerline{\includegraphics[width=0.4\textwidth]{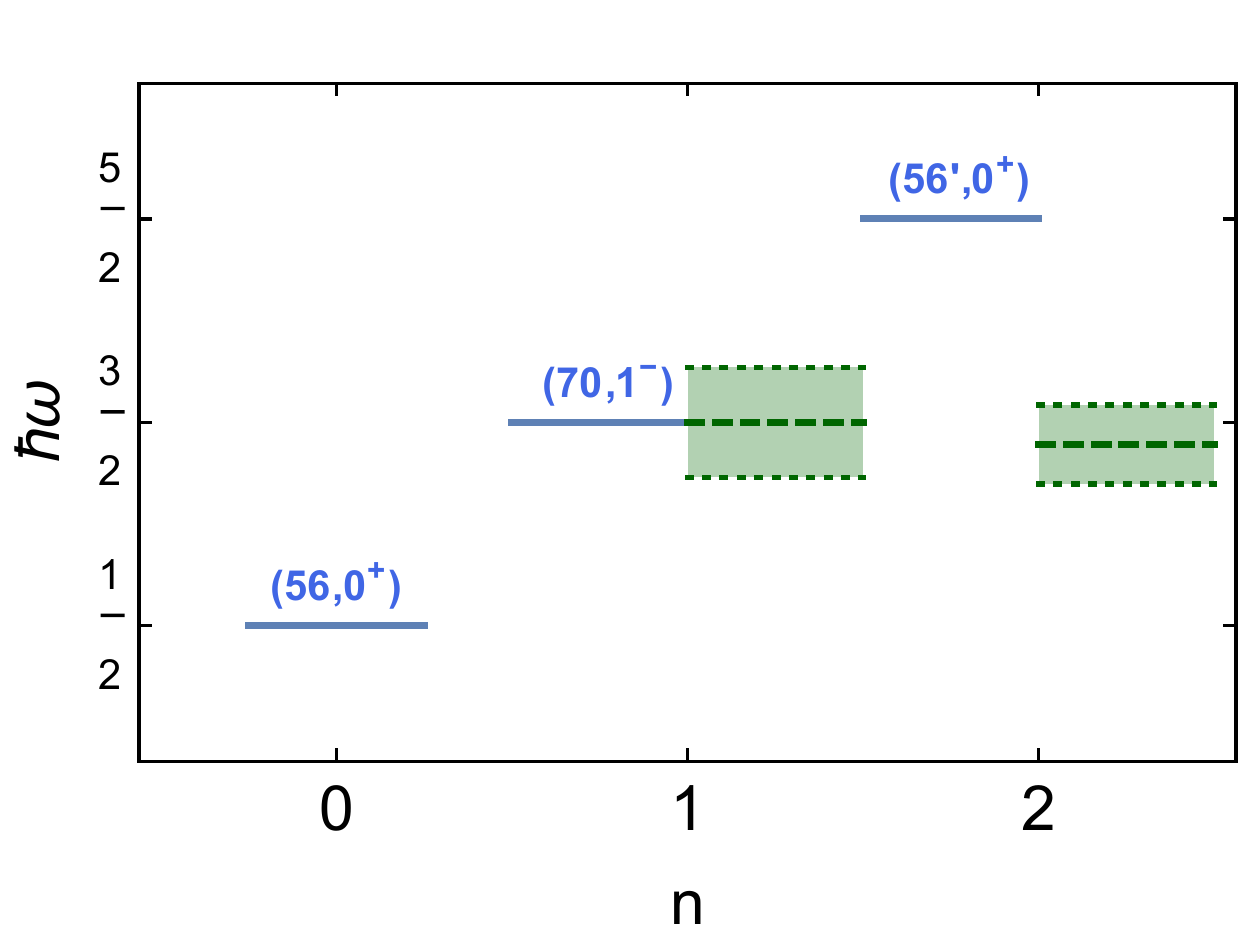}}
\caption{\label{HOlevels}
Blue lines: level ordering produced by the Hamiltonian in Eq.\,\eqref{HOHamiltonian}.  The $(56^\prime,0^+)$ level represents a supermultiplet that is completed by the states in the following representations of $SU(3)\times O(3)$: $(56,2^+)$, $(20,1^+)$, $(70,2^+)$, $(70,0^+)$.
Green dashed lines and shaded bands: pole-mass and width of the nucleon's two lowest-lying $J=1/2$ excitations, determined in a wide ranging analysis of available data \cite{Kamano:2013iva}.
For the purposes of this illustration, $\hbar \omega$ is chosen so that the proton-$N(1535)\,1/2^-$ splitting associates the $N(1535)\,1/2^-$ state with the $(70,1^-)$ supermultiplet, as suggested in quantum mechanics by its spin and parity.
}
\end{figure}

The Roper was found during a dramatic period in the development of hadron physics, which saw the
the appearance of ``color'' as a quantum number carried by ``constituent quarks'' \cite{Greenberg:1964pe}, the interpretation of baryons as bound states of three such constituents \cite{GellMann:1964nj, Zweig:1981pd}, and the development of nonrelativistic quantum mechanical models with two-body potentials between constituent quarks that were tuned to describe the baryon spectrum as it was then known \cite{Hey:1982aj}.  Owing to their mathematical properties, harmonic oscillator potentials were favored as the zeroth-order term in the associated Hamiltonian:
\begin{equation}
\label{HOHamiltonian}
H_0=T + U_0\,, \; T = \sum_{i=1}^3 \frac{p_i^2}{2 M_i} \,, \;
U_0 = \sum_{i<j=1}^3 \tfrac{1}{2} K r_{ij}^2,
\end{equation}
where $p_i$ are the constituent-quark momenta, $r_{ij}$ are the associated two-body separations, and spin-dependent interactions were treated as [perturbative] corrections.  The indices in Eq.\,\eqref{HOHamiltonian} sample the baryon's constituent-quark flavors so that, \emph{e.g}.\ in the proton, $\{1,2,3\}\equiv\{U={\rm Up},U={\rm Up},D={\rm Down}\}$, and $K$ is a common ``spring constant'' for all the constituents.  If one assumes that all three constituent-quarks have the same mass, \emph{viz}.\ $M_1=M_2=M_3$, then this Hamiltonian produces the level ordering in Fig.\,\ref{HOlevels}.  [A similar ordering of these low-lying levels is also obtained with linear two-body potentials \cite{Richard:1992uk}.]

\begin{figure}[t]
\centerline{\includegraphics[width=0.47\textwidth]{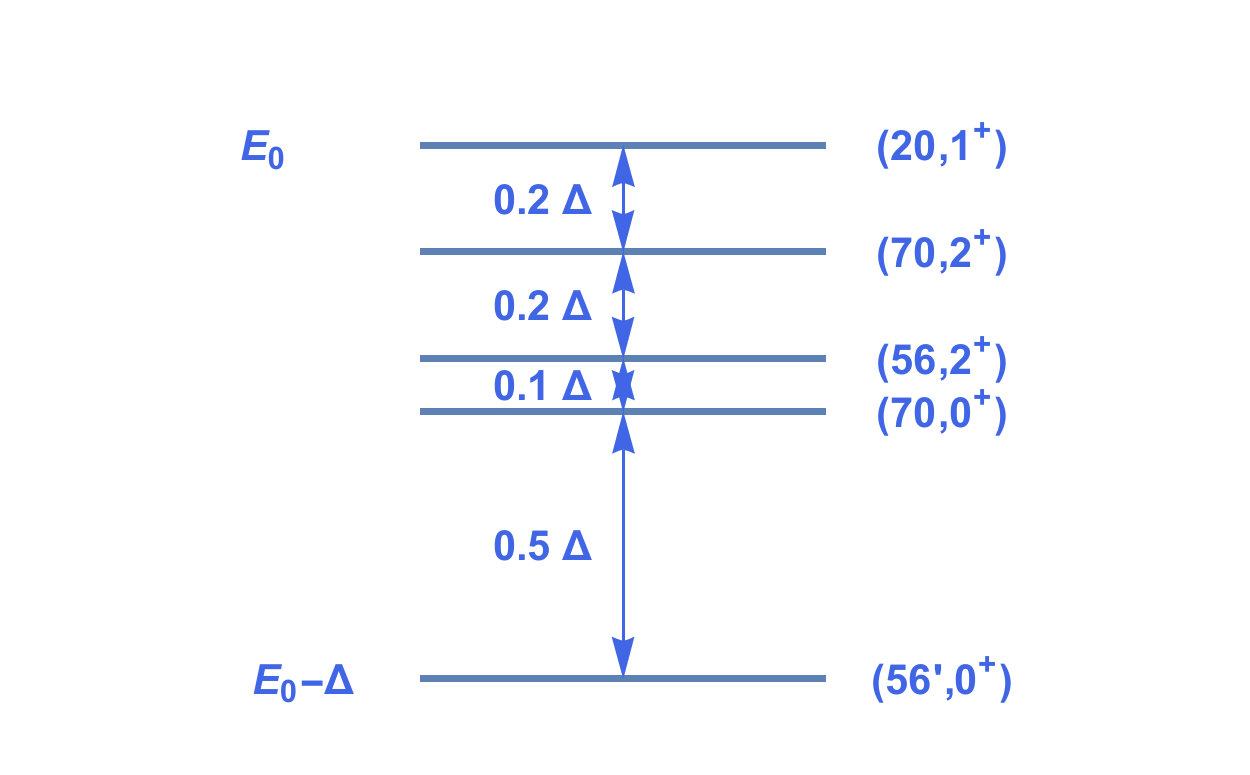}}
\caption{\label{n2splitting}
If an arbitrary anharmonic potential, restricted only insofar as it can be written as the sum of two-body potentials, is added to $H_0$ in Eq.\,\eqref{HOHamiltonian}, then at first order in perturbation theory the $n=2$ harmonic oscillator supermultiplet is split as indicated here.  [$E_0$ is roughly the original $(56^\prime,0^+)$ energy and $\Delta$ is a measure of the shape of the potential].
}
\end{figure}

It is evident in Fig.\,\ref{HOlevels} that the natural level-ordering obtained with such potential models has the first negative-parity $\Delta L=1$ angular momentum excitation of the ground state three-quark system -- the $N(1535)\,1/2^-$ -- at a lower energy than its first radial excitation.  If the Roper resonance, $N(1440)1/2^+$, is identified with that radial excitation, whose quantum numbers it shares, then there is immediately a serious conflict between experiment and theory. However, this ignores the ``perturbations'', \emph{i.e}.\ corrections to $H_0$, which might describe spin-spin, spin-orbit, and other kindred interactions, that can eliminate the degeneracies in $n\geq 2$ harmonic oscillator supermultiplets.  [There are no such degeneracies in the $n=0,1$ supermultiplets.]  In this connection it was proved \cite{Gromes:1976cr, Isgur:1978wd} that given any anharmonic perturbation of the form $\sum_{i<j}U(r_{ij})$, then at first-order in perturbation theory the $n=2$ supermultiplet is always split as depicted in Fig.\,\ref{n2splitting}, where $\Delta$ is a measure of the shape of the potential.  In practice, there is always a value of $\Delta$ for which the $(56^\prime,0^+)$ [Roper] state is shifted below the $N(1535)\,1/2^-$.  Typically, however, the value is so large that one must question the validity of first-order perturbation theory \cite{Isgur:1978wd}.

Notwithstanding such difficulties, at this time it was not uncommon for practitioners to imagine that such models were providing a realistic picture of the baryon spectrum and, in fact, they were a ``\emph{phenomenal phenomenological success}'' \cite{Hey:1982aj}.  Indeed, there was a perception \cite{Hey:1982aj} that: ``\emph{Although there may still be some weakly coupled resonances lurking around the noise level or background of partial-wave analyses, it seems clear that the major features of the spectrum are known. It is not at all clear that we will ever have much more than, at best, a rudimentary outline of charmed or bottom baryon spectroscopy and it is probable that we have now identified over 90\% of the resonant states that we shall ever disentangle from the experimental data.  Indeed, given present experimental trends, it seems probable also that little more, if any, new experimental data relevant to the baryon spectrum will be forthcoming}.''   Such conclusions were premature, as made clear by Sec.\,\ref{Experiment} herein and also the vast array of novel experimental results from the Belle, BaBar, BESIII and LHCb collaborations \cite{BraatenFB21, ChengpingFB21, Aaij:2015tga}, which reveal states that cannot be explained by quark models.

This period of enthusiasm coincided with the ``discovery'' of QCD, \emph{i.e}.\ the high-energy physics community were convinced that a fundamental relativistic quantum field theory of the strong interaction had been found \cite{Marciano:1977su, Marciano:1979wa}.  Some of its peculiar features had been exposed on the perturbative domain \cite{Politzer:2005kc, Wilczek:2005az, Gross:2005kv}, but the spectrum of bound-states it supported could not then be determined.\footnote{
It may still be said today that the complete spectrum of bound states supported by real QCD, \emph{i.e}.\ in the presence of dynamical quarks with realistic values for their current masses, is unknown.
}
In the absence of approaches with a direct QCD connection, studies of quantum mechanical constituent quark models [CQMs] continued; and in relation with the Roper resonance it was found that within a broad class of plausible phenomenological potentials, the negative-parity orbital excitation of the three-quark ground-state is always lower in mass than the $L=0$ radial excitation \cite{Hogaasen:1982rb, Richard:1992uk}.  This means, \emph{e.g}.\  that the ordering in Fig.\,\ref{n2splitting} is an artifact of first-order perturbation theory, which is unreliable when the leading correction is comparable to the value of $\hbar \omega$ associated with $H_0$; and, moreover, that the ordering of the nucleon's low-lying excitations is incorrect in a wide array of such constituent-quark models \cite{Capstick:2000qj, Crede:2013sze, Giannini:2015zia}.

The difficulty in providing a sound theoretical explanation of the Roper resonance was now becoming apparent.  In fact, at this point it was considered plausible that the $N(1440)\,1/2^+$ might not actually be a state generated by three valence quarks.  Instead, the notion was entertained that it may be a hybrid, \emph{viz}.\ a system with a material valence-gluon component or, at least, that the Roper might contain a substantial hybrid component \cite{Barnes:1982fj, Li:1991yba, Capstick:2002wm}. 

The appearance of QCD refocused attention on some prominent weaknesses in the formulation of CQMs.  In particular, their treatment of constituent-quark motion within a hadron as nonrelativistic, when calculations showed $\langle p_i \rangle \sim M_i$, where $\langle p_i\rangle$ is the mean-momentum of a bound constituent-quark; and the use of nonrelativistic dynamics, \emph{e.g}.\ the omission of calculable relativistic corrections to the various potential terms, which would normally become energy-dependent.  Consequently, a relativized constituent-quark model was developed \cite{Godfrey:1985xj} and applied to the baryon spectrum \cite{Capstick:1986bm}; but these improvements did not change the ordering of the energy levels, \emph{i.e}.\ the low-lying excitations of the nucleon were still ordered as depicted in Fig.\,\ref{HOlevels}.
This remains true even within a relativistic field theory framework that employs instantaneous interquark interactions to compute the baryon spectrum \cite{Loring:2001kx}; namely, a three-body term expressing linear confinement of constituent-quarks and a spin-flavor dependent two-body interaction to describe spin-dependent mass splittings.

The QCD-inspired CQMs described above all assume that interquark dynamics derives primarily from gluon-related effects.  An alternative is to suppose that the hyperfine interaction between constituent-quarks is produced by exchange of the lightest pseudoscalar mesons \cite{Glozman:1995fu}, \emph{i.e}.\ the pseudo--Nambu-Goldstone modes: $\pi$-, $K$- and $\eta$-mesons, in which case the hyperfine interaction is flavor-dependent, in contrast to that inferred from one-gluon exchange.  Using straightforward algebraic arguments, one may demonstrate that this sort of Goldstone-boson-exchange (GBE) hyperfine interaction produces more attraction in systems whose wave functions possess higher spin-flavor symmetry.  Such dynamics can thus lead to an inversion of the excited state levels depicted in Fig.\,\ref{HOlevels}, so that the Roper resonance, viewed as the lowest radial excitation of a three constituent-quark ground state, lies below the $N(1535)\,1/2^-$, which is the first orbital excitation of that system \cite{Yang:2017}.  This inversion of levels is a positive feature of the model; and it hints that meson-like correlations should play a role in positioning states in the baryon spectrum.  [Similar conclusions may be drawn from analyses of unquenched CQMs \cite{JuliaDiaz:2006av}.]

On the other hand, a GBE picture of baryon structure can only be figurative, at best.  All mesons are composite systems, with radii that are similar in magnitude to those of baryons; and hence one-boson exchange between constituent-quarks cannot be understood literally.  Moreover, serious difficulties of interpretation arise immediately if one attempts to compute the meson spectrum using a similar Hamiltonian, \emph{e.g}.\ how are the pointlike bosons exchanged between constituent quarks to be understood in the context of the nonpointlike mesonic bound states they help produce?

A deeper class of questions is relevant to all such CQMs.  Namely, in the era of QCD: can any connection be drawn between that underlying theory and the concept of a constituent quark; can the interactions between the lightest quarks in nature veraciously be described by a potential, of any kind; and notwithstanding the challenges they face in describing the Roper resonance, do their apparent successes in other areas yield any sound insights into strong interaction phenomena?  At present, each practitioner has their own answers to these questions.  Our view will subsequently emerge, but is readily stated: used judiciously, CQMs continue to be valuable part of the sQCD toolkit.


\setcounter{equation}{0}
\setcounter{figure}{0}
\section{Roper Resonance in Experiment}
\label{Experiment}
\subsection{Sparse Data}
\begin{figure}[t]
\centerline{\includegraphics[width=0.49\textwidth]{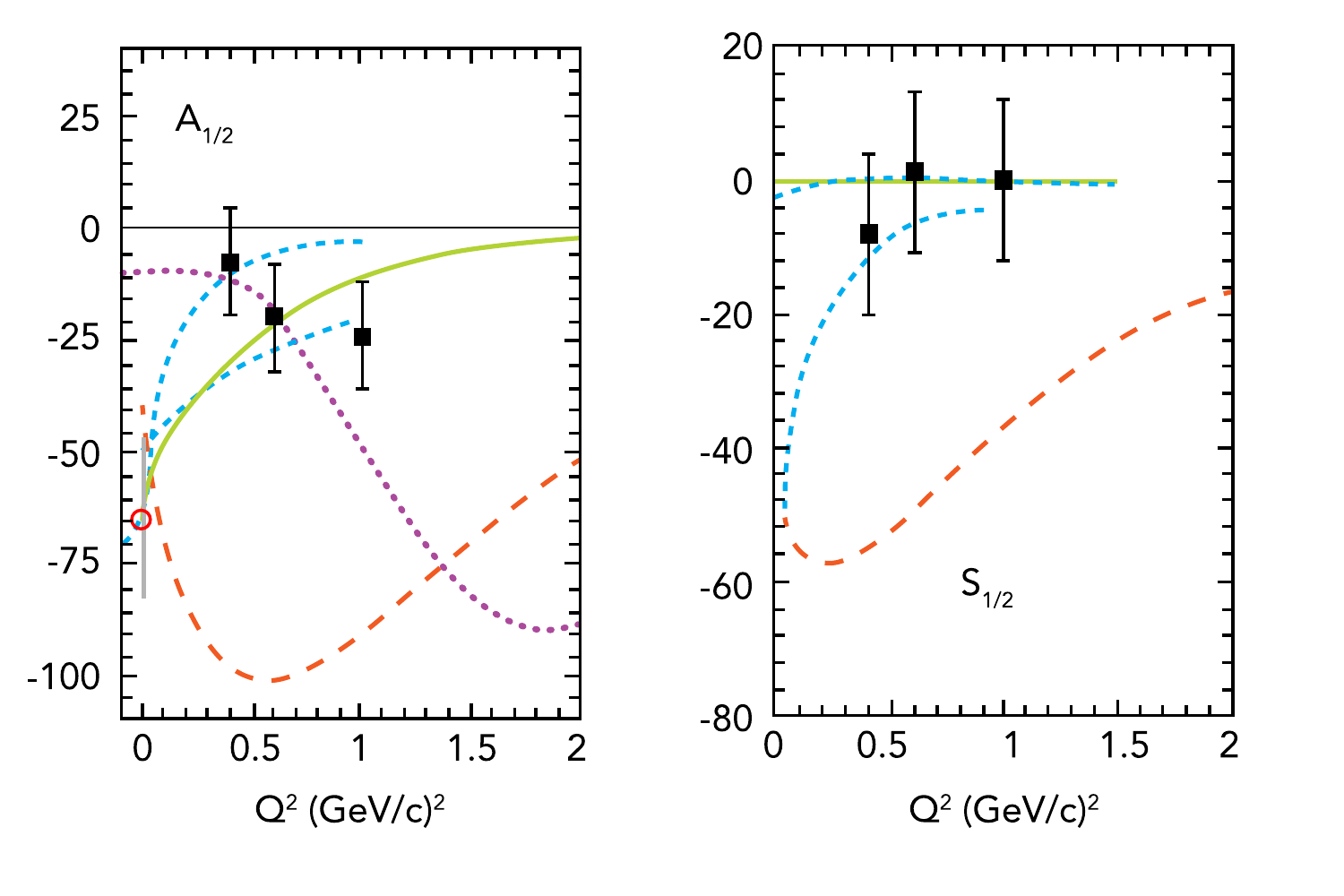}}
\caption{\label{OldElectrocouplings}
Data on the transverse [left panel] and longitudinal [right] photo- and electrocoupling helicity amplitudes for the Roper resonance, Eqs.\,\eqref{ThHelAmp}, as they were available in the last millennium.
%
%
\emph{Legend}.
Data:
open [red] circle -- 1998 estimate of $A_{1/2}$ at the photoproduction point \cite{Caso:1998} and 
error bar [gray] -- our assessment of the true uncertainty in this value at that time;
and solid squares and short-dashed [cyan] curves -- results from a fixed-$t$ dispersion relation fit \cite{Gerhardt:1980yg}, where the error bars on the squares are our estimate of the systematic uncertainty in these values.
Illustrative model results:
long-dashed [red] curves -- non-relativistic quark model \cite{Koniuk:1979vy, Close:1989aj} [incompatible with then-existing data];
dotted curve [purple, left panel] -- relativized quark model \cite{Warns:1989ie};
and solid curve [green] -- model constructed assuming the Roper is a hybrid system, constituted from three constituent-quarks plus a type of gluon excitation \cite{Li:1991yba}, wherewith the longitudinal amplitude vanishes.
[The ordinate is expressed in units of $10^{-3}GeV^{-1/2}$.]}
\end{figure}

%
One material source of the difficulty in understanding the Roper resonance is the quality of the data that was available in the previous millennium.  Illustrated by Fig.\,\ref{OldElectrocouplings}, it was poor owing to limitations in sensitivity to the channels $\gamma p \to \pi^0 p$ and $ep \to e\pi^0 p$ that were typically employed in analyses of the photo- and electrocoupling helicity amplitudes and transition form factors.  Such data could not reasonably be used to distinguish between competing theoretical models of the Roper resonance.  It was therefore evident, given that physics is an empirical science, that a key to resolving the conundrum was more and better data, \emph{i.e}.\ to replace the very limited amount of data available in the previous millennium with a much larger set of high-precision data. This was a strong motivation for a new experimental program at what is now known as the Thomas Jefferson National Accelerator Facility [JLab], which began operations in 1994 and was then called the Continuous Electron Beam Accelerator Facility [CEBAF].

\subsection{Electroproduction Kinematics}
The data in Fig.\,\ref{OldElectrocouplings} were obtained in $e N \to e \pi N$ reactions, \emph{i.e}.\ single-pion photo- and electroproduction processes.
The production of a nucleon resonance in the intermediate part of such reactions is described by the following electromagnetic current, which connects $J=1/2$-baryon initial and final states and is completely expressed by two form factors:
\begin{equation}
\bar u_{f}(P_f)\big[ \gamma_\mu^T F_{1}^{\ast}(Q^2)+\frac{1}{m_{{fi}}} \sigma_{\mu\nu} Q_\nu F_{2}^{\ast}(Q^2)\big] u_{i}(P_i)\,,
\label{NRcurrents}
\end{equation}
where: $u_{i}$, $\bar u_{f}$ are, respectively, Dirac spinors describing the incoming/outgoing baryons, with four-momenta $P_{i,f}$ and masses $m_{i,f}$ so that $P_{i,f}^2=-m_{i,f}^2$; $Q=P_f-P_i$; $m_{{fi}} = (m_f+m_{i})$; and $\gamma^T \cdot Q= 0$.
In terms of these quantities, the helicity amplitudes in Fig.\,\ref{OldElectrocouplings} are:
\begin{subequations}
\label{ThHelAmp}
\begin{align}
A_{\frac{1}{2}}(Q^2) & = c(Q^2) \left[ F_{1}^{\ast}(Q^2)+ F_{2}^{\ast}(Q^2) \right],\\
 S_{\frac{1}{2}}(Q^2) & =  \frac{q_{\rm CMS}}{\surd 2} c(Q^2)
\left[ F_{1}^{\ast}(Q^2) \frac{m_{fi}}{Q^2}  - \frac{F_{2}^{\ast}(Q^2)}{m_{fi}}  \right],
\end{align}
\end{subequations}
with
\begin{equation}
c(Q^2) = \left[ \frac{ \alpha_{\rm em} \pi  Q^2_-}{m_f m_i K} \right]^{\frac{1}{2}},\;
q_{\rm CMS} = \frac{\sqrt{Q_-^2 Q_+^2}}{2 m_f},
\end{equation}
where $Q_{\pm}^2 = Q^2 + (m_f \pm m_i)^2$, $K= (m_f^2-m_i^2)/(2 m_f)$.
%

The dominant Roper decay is $N(1440) \to N\pi$, where the neutron$+\pi^+$ $(n\,\pi^+)$ channel is most prominent.  It also couples to the two-pion channel, being there most conspicuous in $N(1440) \to p\,\pi^+\pi^-$, where $p$ labels the proton.  By design, the CEBAF Large Acceptance Spectrometer [CLAS] at JLab was ideally suited to measuring both these reactions in the same experiment, simultaneously employing the polarized high-precision continuous-wave electron beam at energies up to 6\,GeV.  This capability provided the CLAS Collaboration with a considerable advantage over earlier experiments because measurements and extractions of Roper resonance observables could be based on the analysis of complete centre-of-mass angular distributions and large energy range, and cross-checked against each other in different channels.

\begin{figure}[t]
\centerline{\includegraphics[width=0.42\textwidth]{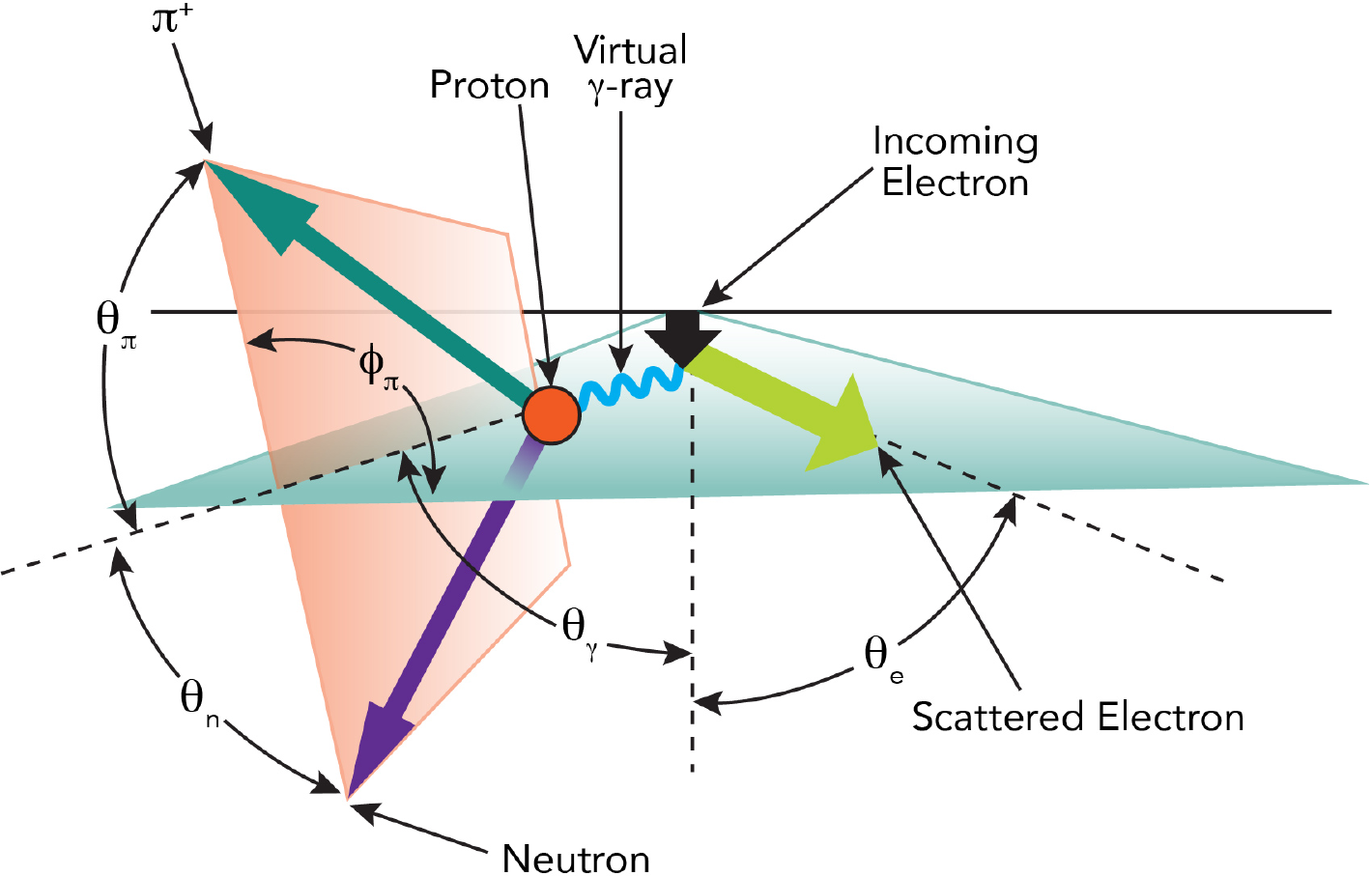}}
\caption{\label{kinematics}
Kinematics of $\pi^+$ electroproduction from a proton.}
\end{figure}

A typical choice of kinematics for the reaction $ep\to en\pi^+$ is depicted in Fig.\,\ref{kinematics}: the incoming and outgoing electrons define the scattering plane; the $\pi^+$ and neutron momentum vectors define the hadronic production plane, characterised by polar angles $\theta_{\pi}$ and $\theta_n$; and the azimuthal angle $\phi_\pi$ defines the angle between the production plane and the electron scattering plane.  In these terms, the differential cross-section can be written:
\begin{equation}
{d^3\sigma \over dE_f d\Omega_e d\Omega} =: {\Gamma {d\sigma \over d\Omega}},
\end{equation}
where $\Gamma$ is the virtual photon flux:
\begin{equation}
\Gamma = {\alpha_{em} \over 2\pi^2Q^2} {(W^2 - m_N^2)E_f \over 2m_NE_i} {1\over 1-\epsilon}\,.
\end{equation}
Here: $\alpha_{\rm em}$ is the fine structure constant and $m_N$ is the nucleon mass; $W$ is the invariant mass of the hadronic final state; $Q^2=-(e_i - e_f)^2$ is the photon virtuality, where $e_i$ and $e_f$ are the four-momentum vectors of the initial and final state electrons, respectively, and $E_i$ and $E_f$ are their respective energies in the laboratory frame; $\epsilon$ is the polarization factor of the virtual photon; and $\Omega_e$ and $\Omega$ are the electron and the pion solid angles.  The unpolarized differential hadronic cross-section has the following $\phi_\pi$ dependence:
\begin{equation}
\label{dsigmadOmega}
 {d\sigma \over d\Omega} = \sigma_{L+T} + \epsilon \sigma_{TT}\cos{2\phi_\pi} + \sqrt{2\epsilon(1+\epsilon)}\sigma_{LT}\cos{\phi_\pi}\,,
\end{equation}
with the $\phi_\pi$-independent term defined as
 $\sigma_{L+T} = \sigma_T + \epsilon \sigma_L$.
As distinct from photoproduction with real photons, the virtual photon in electroproduction has both transverse and longitudinal polarizations; and resolving the associated kinematic dependences reveals additional information about the  production process, especially interference effects.  By measuring the azimuthal dependence of the cross-section in Eq.\,\eqref{dsigmadOmega}, one can isolate the terms that describe transverse-transverse and transverse-longitudinal interference.  

\begin{figure}[t]
\centerline{\includegraphics[width=0.45\textwidth]{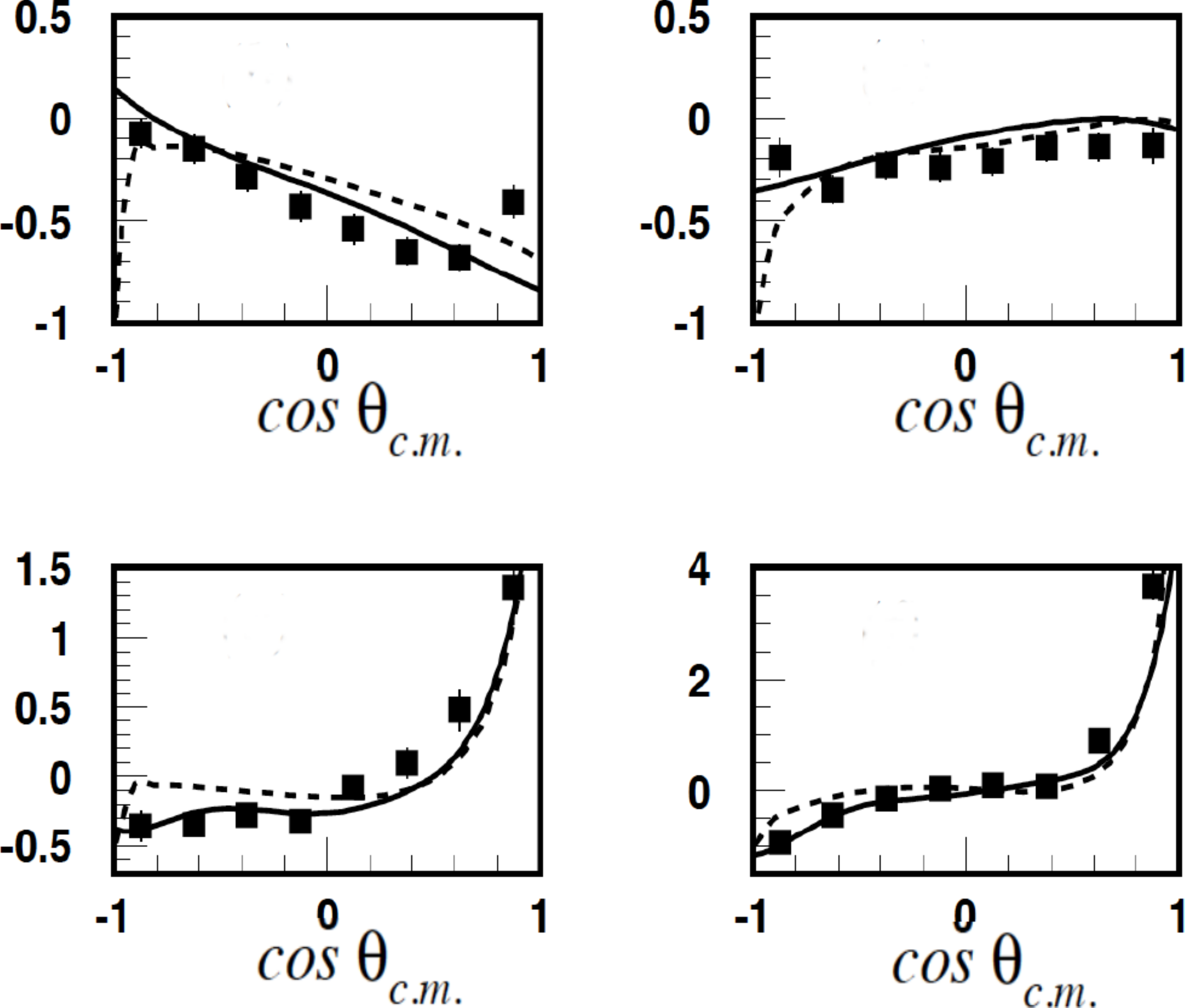}}
\caption{\label{Roper_lowQ}
Cross-section data at $Q^2=0.45\,$GeV$^2$: $\gamma^\ast p \to \pi^0 p$ (upper panels) and $\gamma^\ast p \to \pi^+ n$ (lower panels).  The curves are results of global fits to this data using the UIM [solid] and DR [dashed] approaches.  [Details provided elsewhere \cite{Aznauryan:2004jd}.  The ordinate unit is $\mu$b.]}
\end{figure}

\subsection{Electroproduction Data at Low $\mathbf Q^2$}
Experiments with CLAS began in 1998. Following commissioning, the Collaboration took precise data covering a large mass range from pion threshold up to $W=1.55\,$GeV, \emph{i.e}.\ throughout the first and second resonance regions,\footnote{
The total cross-sections for photo-, electro- and hadro-production of pions from the nucleon exhibit a series of clear ``peak domains'' and each is described as a ``resonance region''.  The first is identified with $W\simeq 1.23\,$GeV (the $\Delta$-baryon); the second, $W\in (1.4,1.6)\,$GeV, contains the Roper resonance, etc.}
with $n\,\pi^+$  and $p\,\pi^0$ final states at two values of $Q^2$, pursuing a primary goal of studying the low-$Q^2$ behavior of the proton-Roper transition.  Analysis of the data was a complex and time-consuming task.

Resonance electroexcitation amplitudes are extracted from exclusive electroproduction data by employing phenomenological reaction models capable of reproducing the full set of observables measured in the $N \pi$ and $p \pi^+ \pi^-$ channels, subject to general reaction theory constraints, such as analyticity and unitarity.  When analysing $n \pi^+$, $p \pi^0$, $p \eta$ final states, the most frequently used approaches are the Unitary Isobar Model (UIM) \cite{Drechsel:1998hk, Aznauryan:2002gd, Drechsel:2007if} and fixed-$t$ dispersion relations \cite{Aznauryan:2004jd}.  In both cases, resonances are described by a relativistic Breit-Wigner distribution involving an energy-dependent width.  Naturally, it is important to implement a good description of the background contributions.  With the UIM approach, these are described explicitly through inclusion of $s$- and $t$-channel meson exchange processes; whereas in the DR method they are calculated directly from the $s$-channel resonance terms using dispersion relations.  The DR approach is tightly constrained, but the UIM method, involving more fitting parameters, has greater flexibility.

\begin{figure}[t]
\centerline{\includegraphics[width=0.42\textwidth]{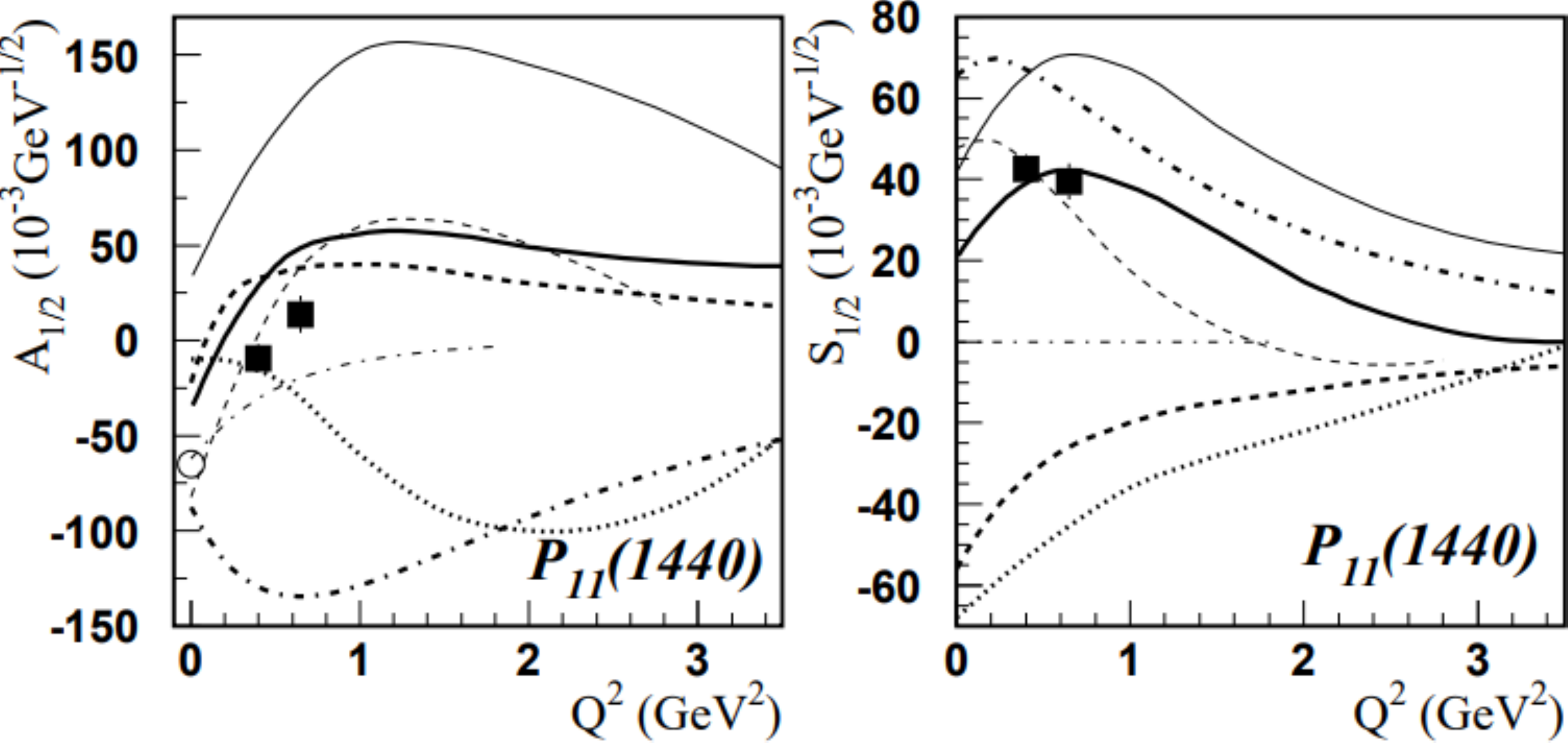}}
\caption{\label{A12_lowQ}
First results from CLAS on the Roper helicity amplitudes \cite{Aznauryan:2004jd} -- solid squares.
All curves are results from various types of CQM:
solid-bold and solid-thin -- results obtained using, respectively, relativistic and non-relativistic versions \cite{Capstick:1994ne};
dotted -- \cite{Warns:1989ie};
dashed \cite{Cano:1998wz};
dot-dashed, thin -- quark-gluon hybrid model \cite{Li:1991yba};
and dot-dashed --  \cite{Tiator:2003uu}.}
\end{figure}

Employing these schemes, the CLAS collaboration released an analysis of their low-$Q^2$ data shortly after the beginning of the new millennium \cite{Aznauryan:2004jd}.   As illustrated by Fig.\,\ref{Roper_lowQ}, both the UIM and DR methods give very similar results; and the Collaboration used the difference between them as an estimate of systematic uncertainties in the model analysis.  In this way they obtained the helicity amplitudes displayed in Fig.\,\ref{A12_lowQ}.  The results contrast starkly with the pre-2000 data in Fig.\,\ref{OldElectrocouplings}: now the transverse amplitude shows a clear zero-crossing near $Q^2=0.5\,$GeV$^2$, the first time this had been seen in any hadron form factor or transition amplitude; and the longitudinal amplitude is large and positive.  The power of precise, accurate data on the transition form factors is also evident in Fig.\,\ref{A12_lowQ}: the hybrid (constituent-quark plus gluon) Roper \cite{Li:1991yba} and two other constituent-quark models \cite{Warns:1989ie, Tiator:2003uu} are eliminated.

The model most favored by the new data is arguably that which describes the Roper as a radial excitation of the nucleon's quark-core dressed by a soft meson cloud \cite{Cano:1998wz}, where a detailed explanation of this ``cloud'' is presented in Sec.\,\ref{sec:DCC}, although the relativistic-CQM \cite{Capstick:1994ne} remains viable.
Notably, both these calculations predict the zero in the $A_{1/2}$ amplitude, although it is achieved through different mechanisms: the meson cloud is responsible in \cite{Cano:1998wz} and relativity plays a crucial role in \cite{Capstick:1994ne}.
It is apparent, too, that the predictions made by these two models are in marked disagreement at larger $Q^2$, \emph{i.e}.\ on the domain within which any soft meson-cloud component of a resonance should become invisible to the probe.  This is correlated with the differing dynamical origins of the $A_{1/2}$ zero in the two CQMs.
It was now clear that higher-$Q^2$ electroproduction data was necessary in order to determine the nature of the Roper resonance.

\begin{figure}[t]
\centerline{\includegraphics[width=0.46\textwidth]{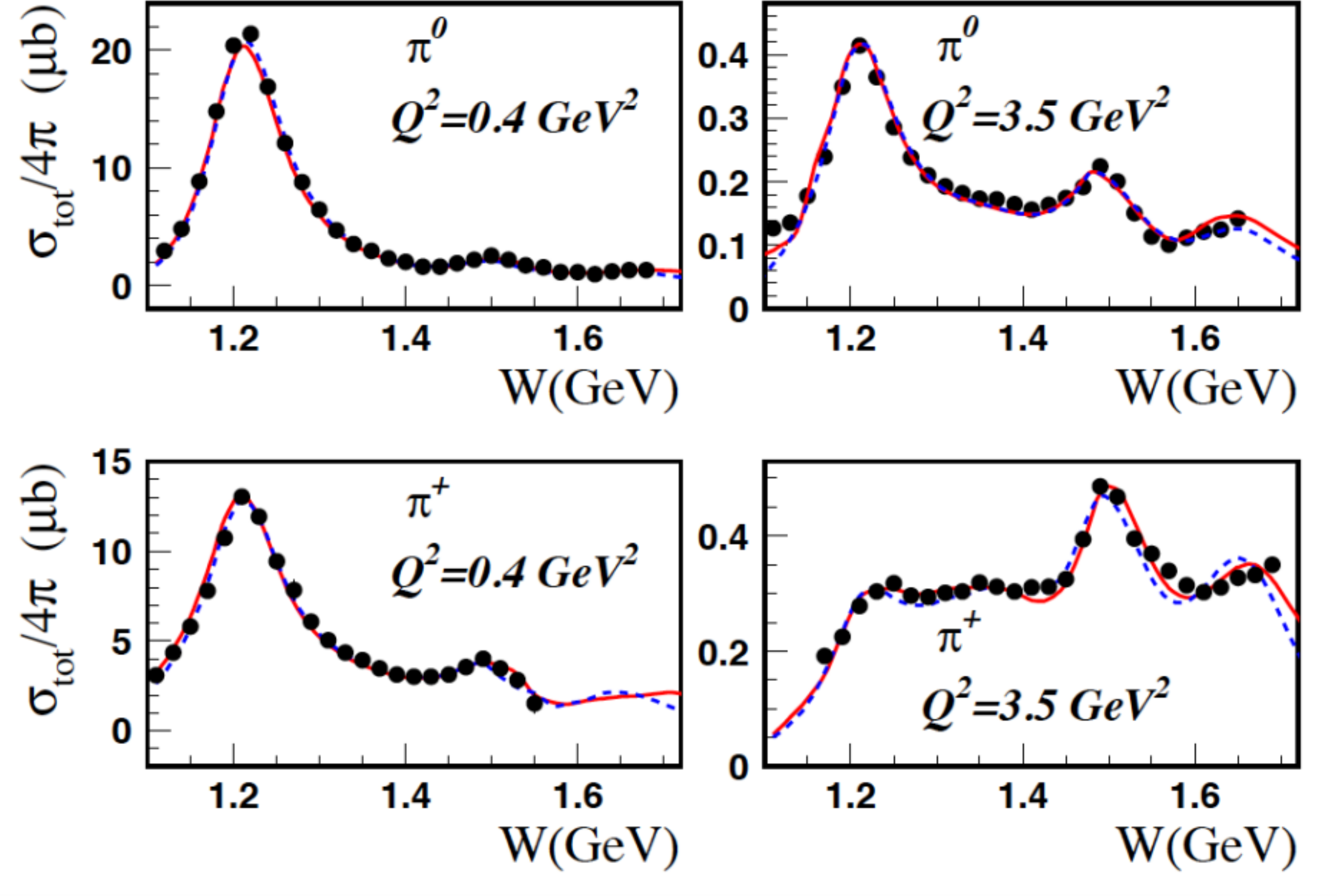}}
\caption{\label{D0_moment}
Lowest moment of the polar-angle dependence in the Legendre expansion of the total cross-section $\sigma_{T+L}$ for the $n\,\pi^+$ and $p\,\pi^0$ electroproduction final states, where the solid [red] and dashed [blue] curves represent, respectively, DR and UIM fits \cite{Aznauryan:2009mx}.  Evidently, whilst the $\Delta(1232)$ is the most conspicuous feature at low-$Q^2$ [left panels], the Roper resonance becomes prominent in the $n \pi^+$ final state at large $Q^2$, generating the broad shoulder centered near $W=1.35\,$GeV [lower right panel].  \emph{N.B}.\ The strong peak at $1.5\,$GeV owes to two other resonances: $N(1520)\,3/2^-$, $N(1535)\,1/2^-$.
}
\end{figure}

\subsection{Pushing electroproduction experiments to higher $\mathbf Q^2$}
Using CLAS and the 6\,GeV continuous-wave electron beam at JLab, high-statistics data were subsequently collected and analyzed, extending the kinematic range to $W=2\,$GeV and $Q^2=4.5\,$GeV$^2$ \cite{Aznauryan:2008pe, Aznauryan:2009mx, Aznauryan:2011qj, Mokeev:2012vsa, Mokeev:2015lda}.
The new experiments revealed some surprising aspects of the Roper electroproduction amplitudes, overturning conclusions that might have been drawn from the low-$Q^2$ data alone.  For example, as highlighted by Fig.\,\ref{D0_moment}, whereas $A_{1/2}$ is small in the low-$Q^2$ range accessed by the earlier CLAS data, because it is undergoing a sign change at $Q^2\approx 0.5\,$GeV$^2$, and hence the Roper is not directly visible in the total cross-section, at high-$Q^2$ this resonance becomes very strong, even dominating over the $\Delta(1232)$ on $Q^2> 2\,$GeV$^2$ in the $n\,\pi^+$ final state.

\begin{figure}[!t]
\centerline{%
\includegraphics[width=0.8\linewidth]{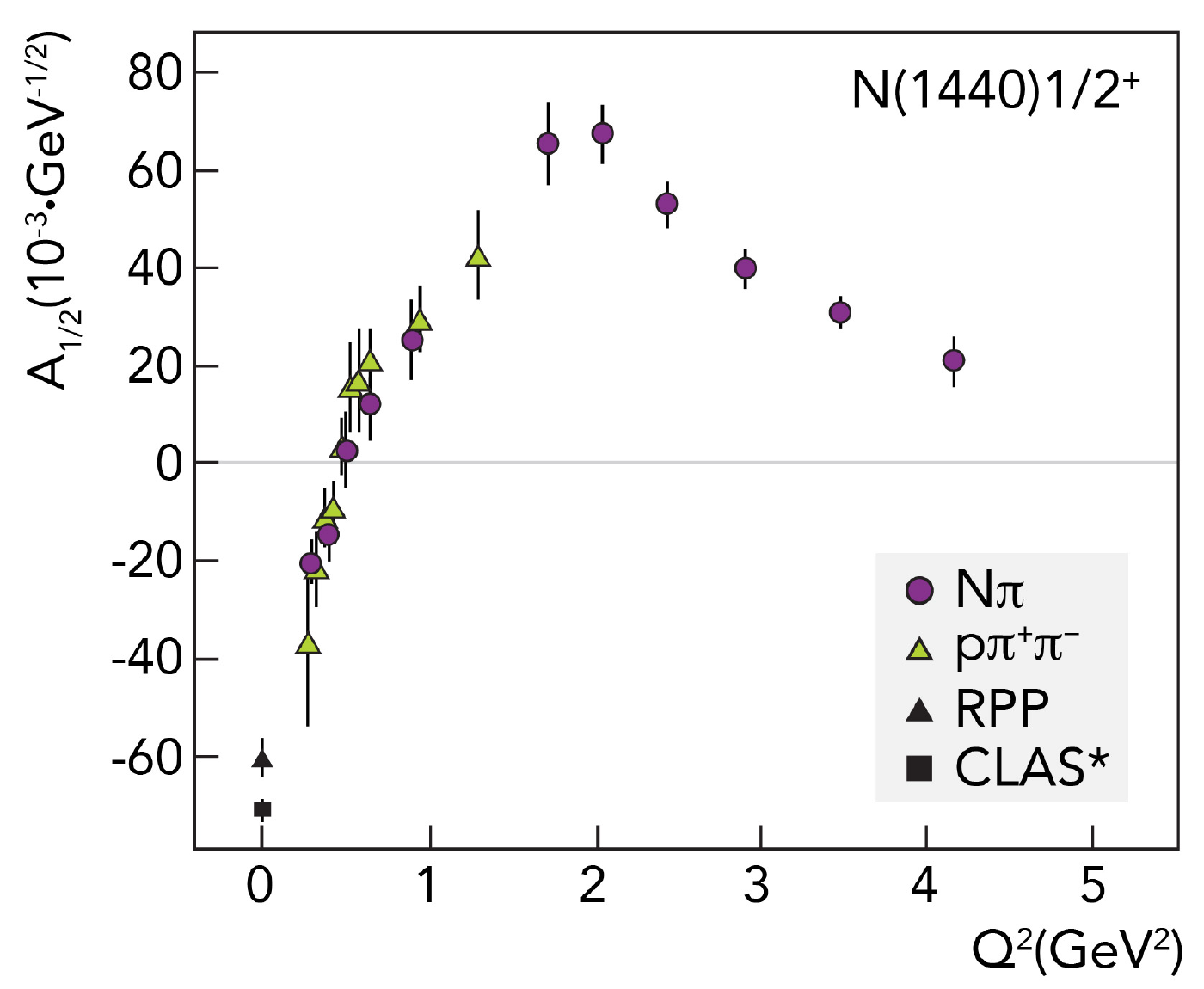}}

\centerline{%
\includegraphics[width=0.8\linewidth]{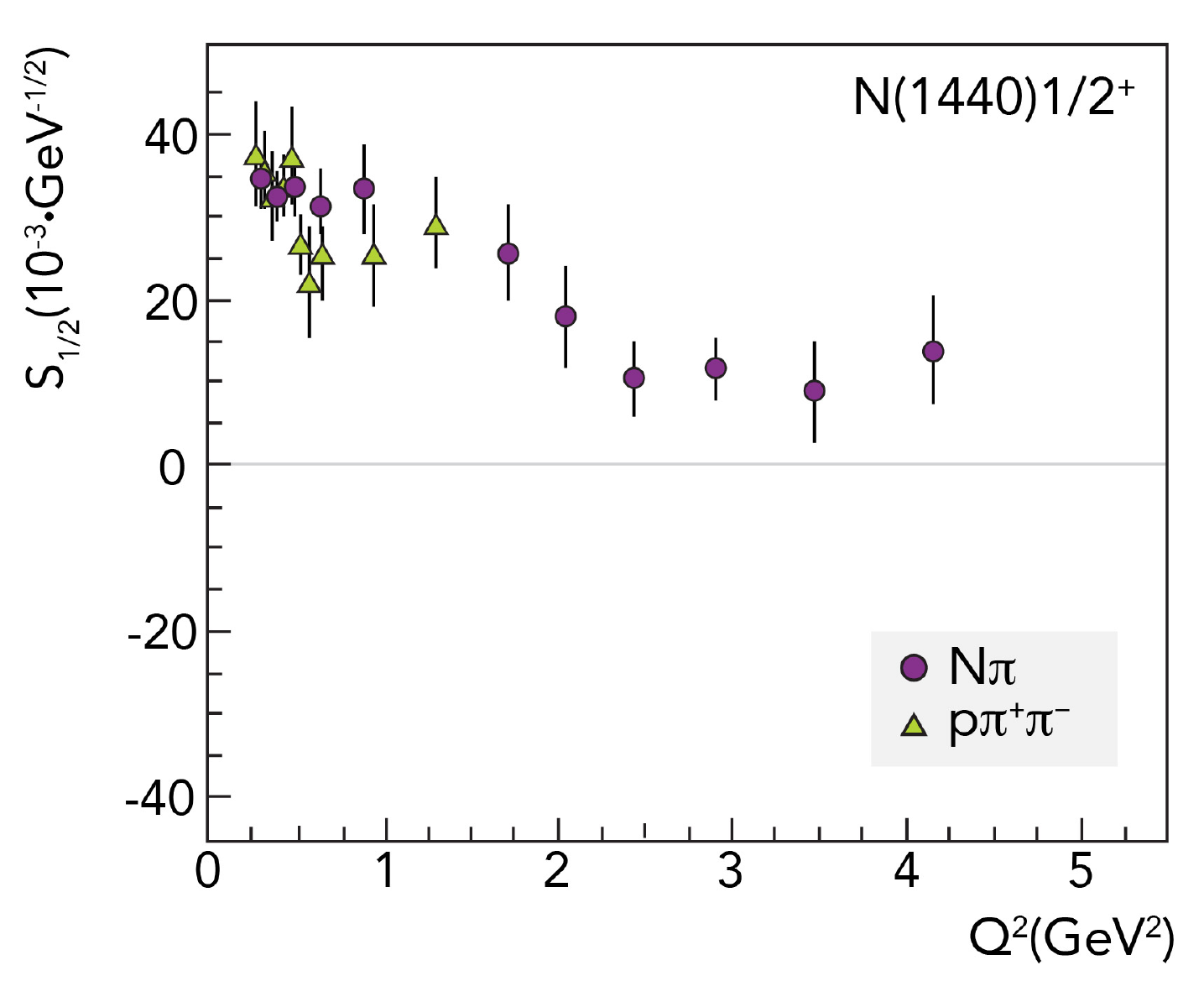}}
%
\caption{\label{A12S12}
Transverse [upper panel] and longitudinal [lower] Roper resonance electrocoupling helicity amplitudes.
Legend:
circles [blue] -- analysis of single-pion final states \cite{Aznauryan:2008pe, Aznauryan:2009mx};
triangles [green] -- analysis of $ep\to e^\prime \pi^+ \pi^- p^\prime$ \cite{Mokeev:2012vsa, Mokeev:2015lda};
square [black] -- CLAS Collaboration result at the photoproduction point \cite{Dugger:2009pn} and
triangle [black] -- global average of this value \cite{Olive:2016xmw}.
}
\end{figure}

The final data set used in the global fit contained over 120\,000 points in $e p \to e^\prime n \pi^+$ and $e p \to e^\prime p \pi^0$, measuring differential cross-section, and polarized beam and polarized target asymmetries, covering the complete range of azimuthal and polar angles, and the domains $W<1.8\,$GeV and $Q^2< 4.5\,$GeV$^2$.  The transverse and longitudinal helicity amplitudes for Roper-resonance electroproduction obtained from the complete analysis are displayed in Fig.\,\ref{A12S12}.  These results confirm those obtained in earlier analyses of much reduced data sets and significantly extend them.  Importantly, the evident agreement between independent analyses of single- and double-pion final states boosts confidence in both.  [\emph{N.B}.\ New CLAS data on $\pi^+ \pi^- p$ electroproduction \cite{Isupov:2017lnd}, with nine one-fold differential cross-sections covering a final hadron invariant mass range $W\in[1.4,2.0]\,$GeV and $Q^2\in [2,5]\,$GeV$^2$, will enable this agreement to be tested further.]



\subsection{Roper Resonance: Current Experimental Status}
It is appropriate here to summarize the modern empirical status.\\[-4ex]
\begin{itemize}
\setlength\itemsep{0em}
\item The Roper [N(1440)\,$1/2^+$] is a four-star resonance with pole mass $\approx 1.37\,$GeV and width $\approx 0.18\,$GeV \cite{Olive:2016xmw}.
\item Transverse helicity amplitude, $A_{1/2}(Q^2)$:\\[-1.7em]
\begin{itemize}
\setlength\itemsep{0em}
\item increases rapidly as $Q^2$ increases from the real photon point to $Q^2 \approx 2\,$GeV$^2$;
\item changes sign at $Q^2 \approx 0.5\,$GeV$^2$;
\item exhibits a maximum value at $Q^2 \approx 2\,$GeV$^2$, attaining a magnitude which matches or exceeds that at the real photon point;
\item decreases steadily toward zero with increasing $Q^2$ after reaching its maximum value.
\end{itemize}
\item Longitudinal helicity amplitude, $S_{1/2}(Q^2)$:\\[-1.7em]
\begin{itemize}
\setlength\itemsep{0em}
\item maximal near the real photon point;
\item decreases slowly as $Q^2$ increases toward 1\,GeV$^2$;
\item decreases more quickly on $Q^2 \gtrsim 1\,$GeV$^2$.
\end{itemize}
\item $N \pi$ and $p \pi^+ \pi^-$ final states in electroproduction:\\
The non-resonant contributions to these two final states are markedly dissimilar and hence very different analysis procedures are required to isolate the resonant contributions.  Notwithstanding this, the results for the resonant contributions agree on the domain of overlap, \emph{i.e}.\ $Q^2 \in [0.25,1.5]\,$GeV$^2$.

\end{itemize} 

\setcounter{equation}{0}
\setcounter{figure}{0}
\section{Dynamical Coupled Channels Calculations}
\label{sec:DCC}
As highlighted in Sec.\,\ref{Experiment}, the last twenty years have seen an explosion in the amount of available data on resonance photo- and electroproduction, \emph{e.g}.\ the reactions $\gamma^{(\ast)} N \to \pi N$ and $\gamma^{(\ast)} N \to \pi \pi N$, which are particularly relevant to discussions of the Roper resonance.  As the data accumulated, so grew an appreciation of the need for a sound theoretical analysis which unified all its reliable elements.  At the beginning of 2006, this culminated with establishment of the Excited Baryon Analysis Center [EBAC] at JLab \cite{Lee:2007zy, Kamano:2011ut, Lee:2012mea}, whose primary goals were: to perform a dynamical coupled-channels [DCC] analysis of the world's data on meson production reactions from the nucleon in order to determine the meson-baryon partial-wave amplitudes; and identify and characterise all nucleon resonances that contribute to these reactions.

In contrast to the familiar and commonly used partial wave analyses, which are model-independent to some extent, but also, therefore, limited in the amount of information they can provide about resonance structure, modern DCC analyses are formulated via a Hamiltonian approach to multichannel reactions \cite{JuliaDiaz:2007kz, Suzuki:2009nj, Kamano:2010ud, Ronchen:2012eg, Kamano:2013iva}.  The Hamiltonian expresses model assumptions, \emph{e.g}.\ statements about the masses of bare/undressed baryons [in the sense of particle versus quasi-particle] and the dominant meson-baryon reaction channels that transform the bare baryon into the observed quasi-particle.  Naturally, such assumptions can be wrong.  Equally: the models are flexible; they can be falsified and thereby improved, given the vast amount of existing data; and, used judiciously, they can be provide a critical bridge between data and QCD-connected approaches to the computation of baryon properties.

The EBAC approach,\footnote{The EBAC projected terminated in 2012, but the effort is continuing as part of the  Argonne-Osaka collaboration, from which it initially grew \cite{Sato:1996gk, Matsuyama:2006rp}.} for instance, describes meson-baryon ($MB$) reactions involving the following channels: $\pi N$, $\eta N$ and $\pi\pi N$, the last of which has $\pi \Delta$, $\rho N$ and $\sigma N$ resonant components.  The excitation of the internal structure of a given initial-state baryon ($B$) by a meson ($M$) to produce a bare nucleon resonance, $\bar N^\ast$, is implemented by an interaction vertex, $\Gamma_{MB\to \bar N^\ast}$.  Importantly, the Hamiltonian also contains energy-independent meson-exchange terms, $v_{MB,M^\prime B^\prime}$, deduced from widely-used meson-exchange models of $\pi N$ and $NN$ scattering.

In such an approach, the features of a given partial wave amplitude may be connected with dressing of the bare resonances included in the Hamiltonian ($\bar N^\ast$), in which case the resulting $N^\ast$ states are considered to be true resonance excitations of the initial state baryon.  On the other hand, they can also be generated by attraction produced by the $v_{MB,M^\prime B^\prime}$ interaction and channel-coupling effects, in which case they are commonly described as ``molecular states'' so as to differentiate them from true resonance excitations.  The need to reliably distinguish between these two different types of systems in the solution of the coupled channels problem defined by the model Hamiltonian requires that the form and features of $v_{MB,M^\prime B^\prime}$ must be very carefully constrained by,  \emph{e.g}.\ elastic scattering data, throughout the region of relevance to the resonance production reactions.

\begin{figure}[t]
\centerline{\includegraphics[width=0.4\textwidth]{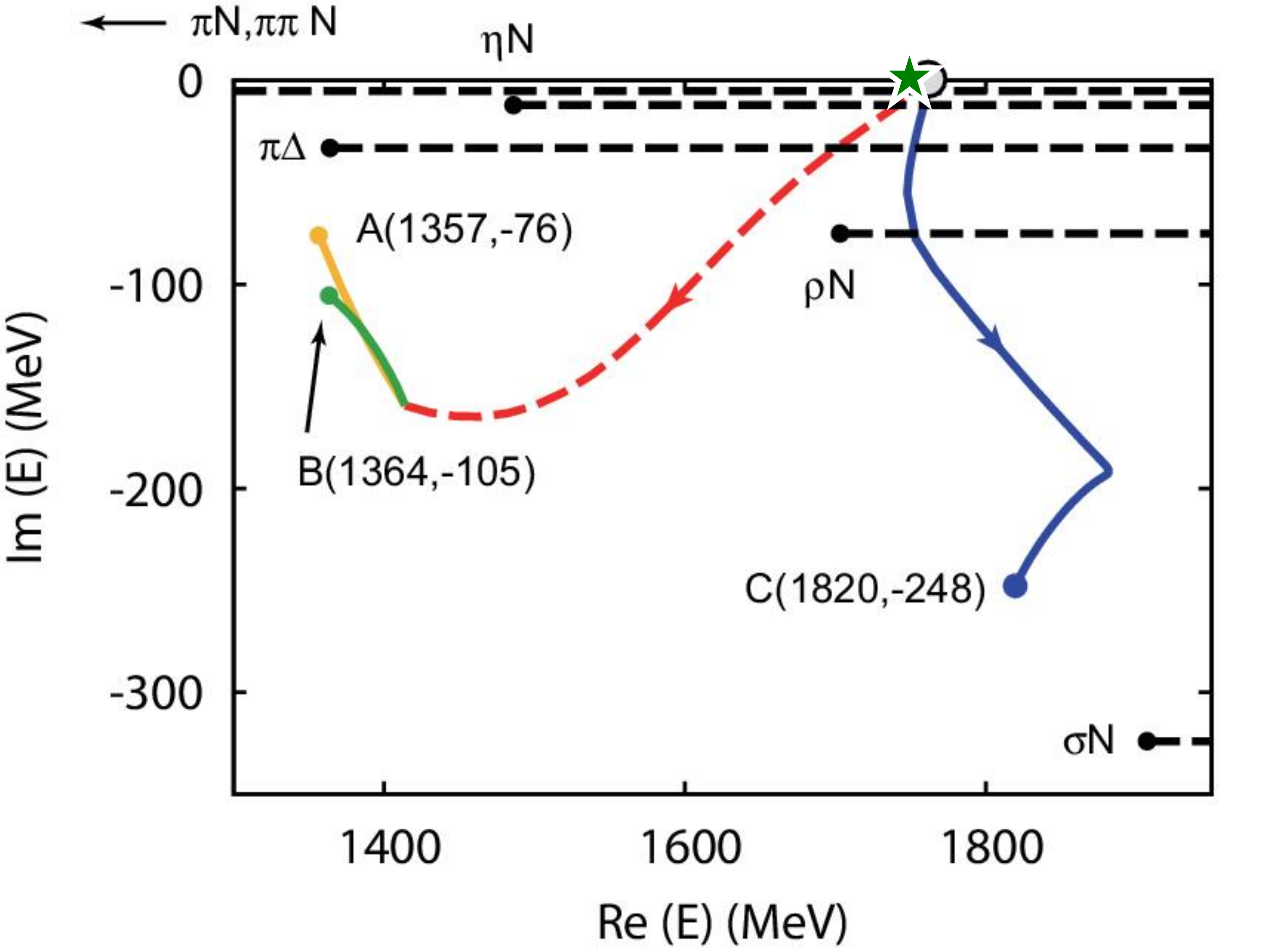}}
\caption{\label{EBACRoper}
Open circle [black]: mass of the bare Roper state determined in the EBAC DCC analysis of $\pi N$ scattering \cite{JuliaDiaz:2007kz, Suzuki:2009nj, Kamano:2010ud}.
This bare Roper state, with full spectral weight at mass $1.763\,$GeV, splits and evolves following the inclusion of meson-baryon final-state interactions, with the trajectories in this complex-energy plane depicting the motion of the three, distinct daughter poles as the magnitude of those interactions is increased from zero to their full strength.
The horizontal dashed lines [black] mark the branch cuts associated with all thresholds relevant to the solution of the DCC scattering problem in this channel.
Filled star [green]: mass of the dressed-quark core of the proton's first radial excitation predicted by a three valence-quark Faddeev equation \cite{Segovia:2015hra}.
}
\end{figure}

Being aware of the challenges associated with understanding the Roper resonance, the EBAC collaboration made a determined effort to produce a sound description of the spectrum of baryon resonances with masses below 2\,GeV using their DCC model.  Refining this tool by developing an excellent description of 22\,348 independent data points, representing the complete array of partial waves, they arrived at some very striking conclusions \cite{JuliaDiaz:2007kz, Suzuki:2009nj, Kamano:2010ud}, illustrated in Fig.\,\ref{EBACRoper}:\\[-4ex]
\begin{itemize}
\setlength\itemsep{0em}
\item From a bare state with mass $1.763\,$GeV, three distinct features appear in the $P_{11}$ partial wave, as described by Fig.\,\ref{EBACRoper}.  [We will subsequently return to the interpretation of the bare state.]
\item Of the three spectral features that emerge in this channel, two are associated with the Roper resonance.  [This two-pole character of the Roper is common to many analyses of the scattering data, including one involving Roper himself \cite{Arndt:1985vj} and more recent analyses of $\pi N$ scattering data \cite{Cutkosky:1990zh, Arndt:2006bf, Doring:2009yv}.]
\item The third pole is located farther from the origin [position C in Fig.\,\ref{EBACRoper}] and might plausibly be associated with the $N(1710)\,1/2^+$ state listed by the Particle Data Group \cite{Olive:2016xmw}.
%

\end{itemize}
[\emph{N.B}.\ The same EBAC DCC analysis identifies a bare state with mass $1.800\,$GeV as the origin of the $N(1535)\,1/2^-$ and a bare state with mass $1.391\,$GeV associated with the $\Delta(1232)\,3/2^+$ \cite{JuliaDiaz:2007kz}.]

Evidently, as emphasized by the trajectories in Fig.\,\ref{EBACRoper}, the coupling between channels required to simultaneously describe all partial waves has an extraordinary effect with, \emph{e.g}.\ numerous spectral features in the $P_{11}$ channel evolving from a single bare state, expressed as a pole on the real axis, through its coupling to the $\pi N$, $\eta N$ and $\pi\pi N$ reaction channels.  It follows that no analysis of one partial wave in isolation can reasonably be claimed to provide an understanding of such a complex array of emergent features.


\setcounter{equation}{0}
\setcounter{figure}{0}
%
\section{Relativistic Quantum Field Theory}
\label{TheorySection}
\subsection{Lattice-regularized QCD}
\label{sec:lQCD}
An introduction to the numerical simulation of lattice-regularized QCD (lQCD) is provided elsewhere \cite{Gattringer:2010zz}; so here we simply note that this method is a nonperturbative approach to solving QCD in which the gluon and quark fields are quantized on a discrete lattice of finite extent, whose intersections each represent a point in spacetime \cite{Wilson:1974sk}.

The lQCD approach has provided a spectrum of light ground-state hadrons that agrees with experiment \cite{Durr:2008zz}, but numerous hurdles are encountered in attempting to compute properties of resonance states in this way \cite{Liu:2016rwa, Briceno:2017max}.
In connection with the Roper, which in reality couples strongly to many final-state interaction [FSI] channels, as indicated in Fig.\,\ref{EBACRoper}, these include the following:
the challenges of computing with a realistic pion mass and developing both a fully-representative collection of interpolating fields and a valid strategy for handling all contributing final-state interaction channels, which incorporate the issue of ensuring that the nucleon's lowest excitations are properly isolated from all higher excitations; and the problem of veraciously expressing chiral symmetry and the pattern by which it is broken in both the fermion action and the algorithm used in performing the simulation.%
\footnote{QCD is asymptotically free \cite{Politzer:2005kc, Gross:2005kv, Wilczek:2005az}.  It is therefore possible to define it in the absence of a quark Lagrangian mass, \emph{viz}.\ a massless theory.  In a truly massless theory, fermions are characterized by their helicity [left or right], no interactions can distinguish between left-handed and right-handed fields, and the theory is therefore chirally symmetric.  In QCD, however, dynamical effects act to destroy this symmetry.}

\begin{figure}[t]


\centerline{
\includegraphics[clip,width=0.42\textwidth]{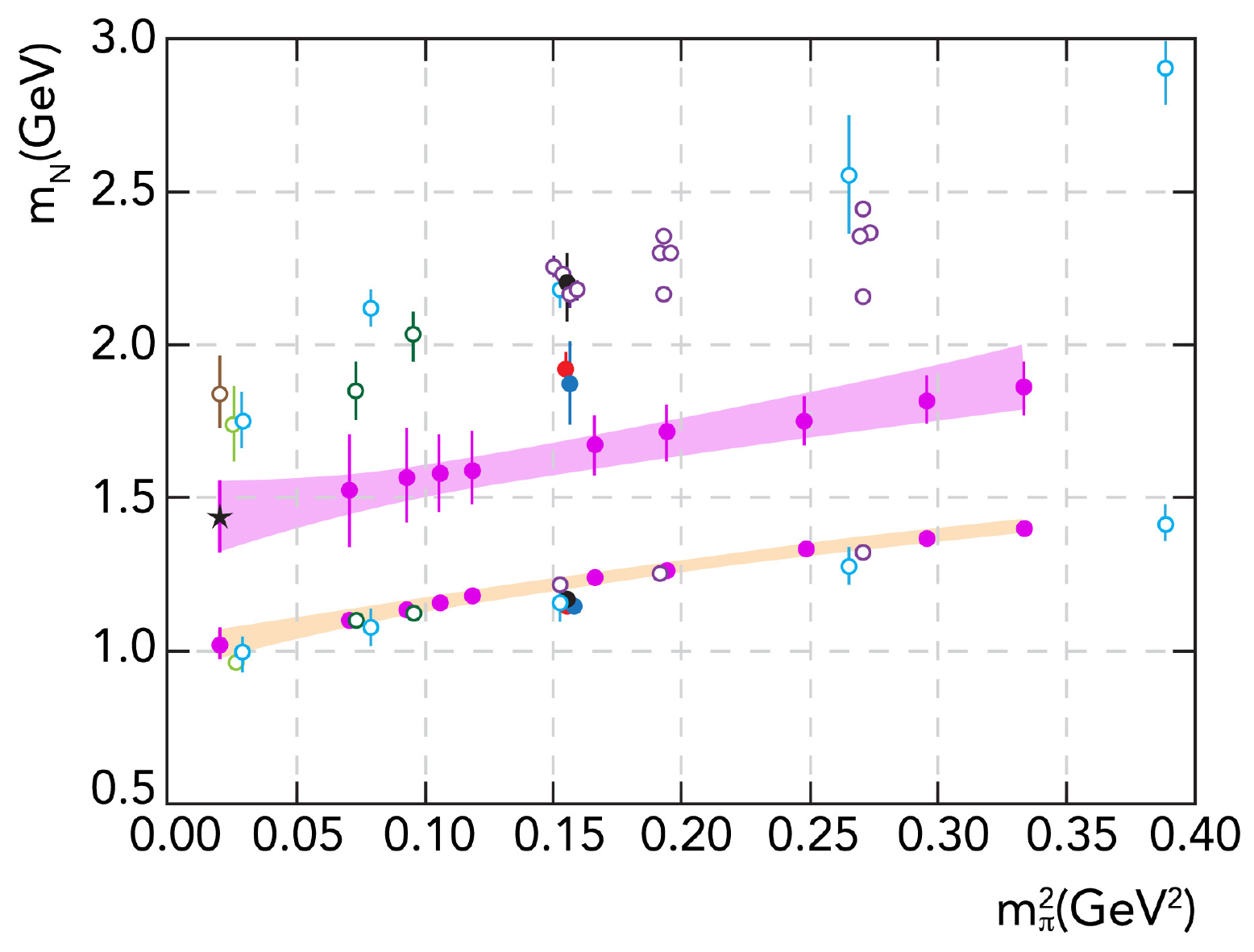}}

\caption{\label{lQCDRoper1}
Illustrative collection of lQCD results for the mass of the nucleon (lower band) and its lightest positive-parity excitation as a function of $m_\pi^2$, where $m_\pi$ is the pion mass used in the simulation.  The results depicted were obtained with different lattice formulations and varying methods for identifying the excited state, as described in the source material \cite{Mahbub:2010rm, Edwards:2011jj, Engel:2013ig, Liu:2014jua, Alexandrou:2014mka} and, in particular, \cite{Liu:2016rwa}.}
\end{figure}

Much needs to be learnt and implemented before these problems are overcome, so the current status of lQCD results for the Roper is unsettled.  This is illustrated in Fig.\,\ref{lQCDRoper1}, which provides a snapshot of recent results for the masses of the nucleon and its lowest-mass positive-parity excitation.  In this image, almost all formulations of the lQCD problem produce values that extrapolate [as $m_\pi^2$ is taken toward its empirical value] to a Roper mass of roughly $1.8\,$GeV, \emph{i.e}.\ to a mass that is 0.4\,GeV above the real part of the empirical value, \emph{viz}. $1.4\,$GeV.  However, one band appears to extrapolate to somewhere near this empirical value.  Contrary to the other formulations, the fermion action in that case \cite{Liu:2014jua} possesses good chiral symmetry properties; and its proponents argue \cite{Liu:2016rwa} that this feature enables the simulation to better incorporate aspects of the extensive dynamical channel couplings which are known to be important in explaining and understanding the spectral features of $\pi N$ scattering in the $P_{11}$ channel \cite{JuliaDiaz:2007kz, Suzuki:2009nj, Kamano:2010ud}.

\begin{figure}[!t]

\centerline{%
\includegraphics[width=0.84\linewidth]{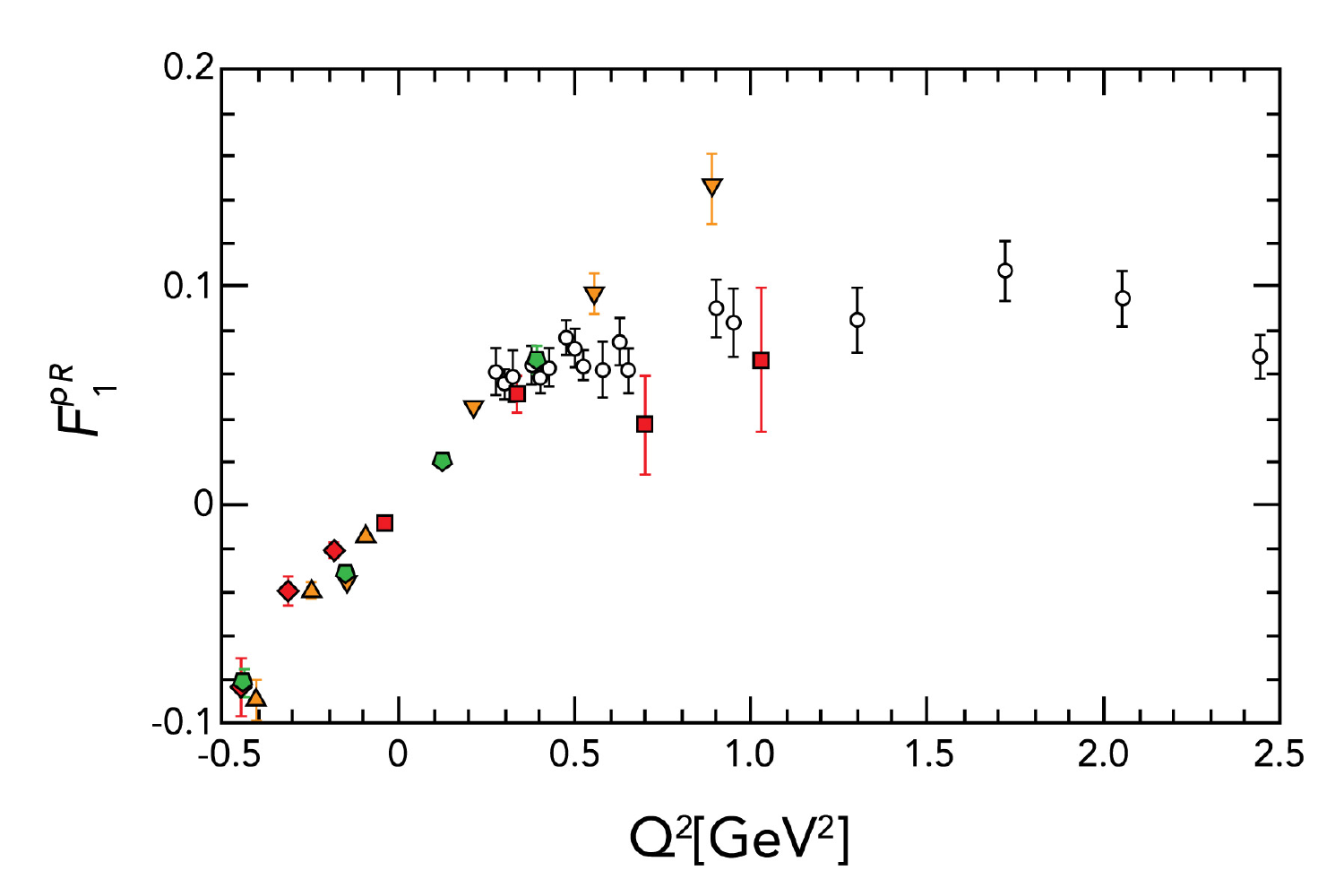}}
\vspace*{1ex}

\centerline{%
\includegraphics[width=0.84\linewidth]{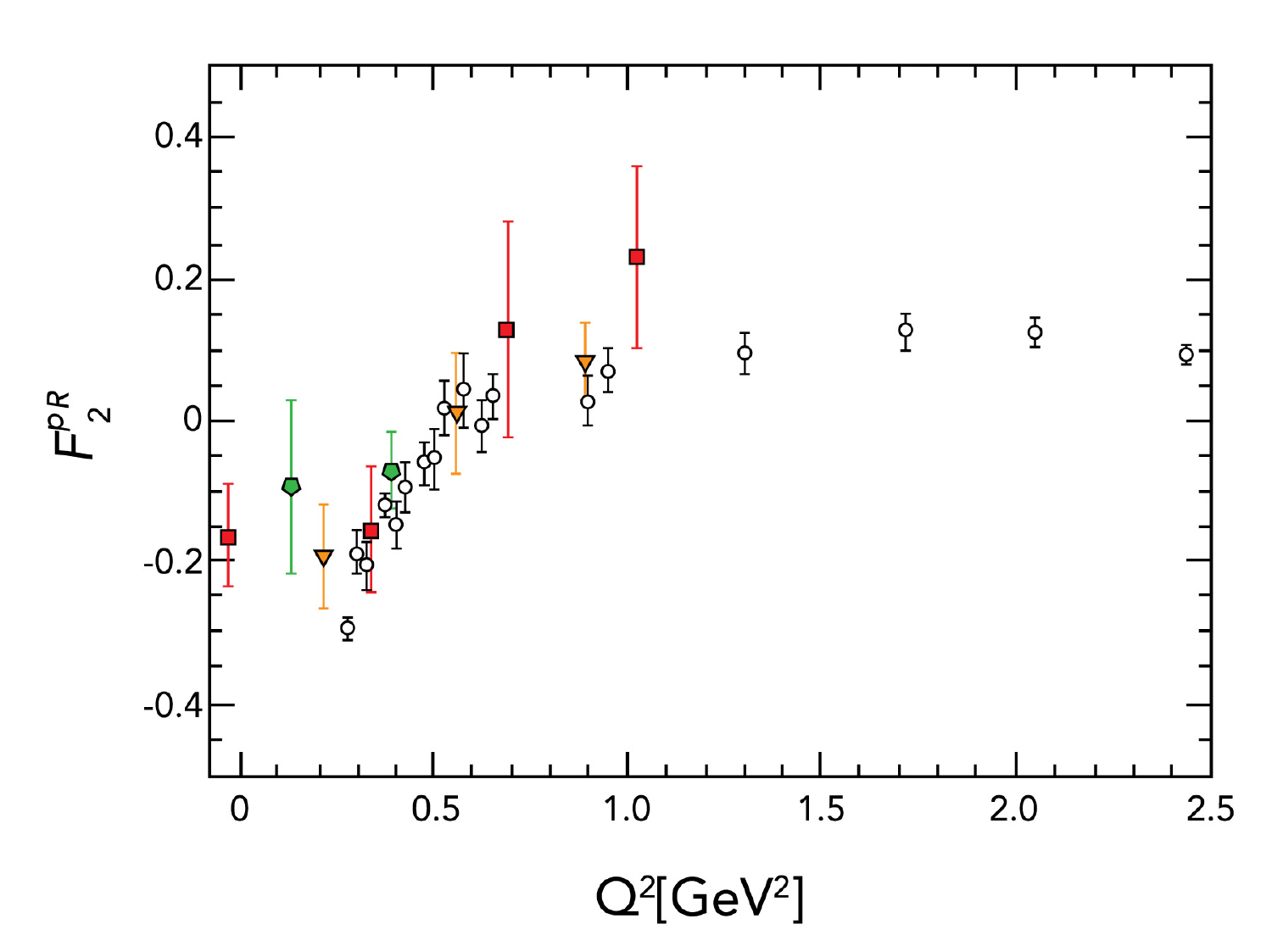}}
%
\caption{\label{lQCDelectroRoper}
Existing results for the Dirac [upper panel] and Pauli [lower] proton-Roper transition form factors computed using the methods of lQCD \cite{Lin:2011da} on anisotropic lattices with pion masses [in GeV]: $0.39$ [red squares], $0.45$ [orange triangles], $0.875\,$ [green circles]; and associated spatial lengths of $3$, $2.5$, $2.5\,$fm.
Open circles are empirical results from the CLAS Collaboration \cite{Aznauryan:2009mx, Dugger:2009pn, Mokeev:2012vsa, Mokeev:2015lda}. }
\end{figure}

As we have emphasized heretofore, computing a value [even correct] for the Roper mass is insufficient to validate a formulation of the Roper resonance problem and its solution.  An additional and far more stringent test is an explanation of the pointwise behavior of the transition form factors measured in electroproduction, Eq.\,\eqref{NRcurrents}.  The first such lQCD calculations, which used the quenched truncation of the theory, are described in \cite{Lin:2008qv}.   More recently, results were obtained with two light quarks and one strange quark [$N_f = 2 + 1$] \cite{Lin:2011da}.  They are depicted in Fig.\,\ref{lQCDelectroRoper}.  These simulations identified the Roper resonance with the first positive-parity excitation of the nucleon, whose computed mass is roughly 1.8\,GeV, and focused on the low-$Q^2$ domain.  Significantly, compared with the quenched results, the inclusion of $N_f = 2 + 1$ dynamical fermions produces a sign change in $F_2^\ast$, located in the same neighborhood as that seen in experimental data.  This difference between quenched and dynamical simulations once again suggests that meson-baryon (MB)\,FSIs are a critical part of the long-wavelength structure of the Roper.

\subsection{Insights from Continuum Analyses}
\label{sec:continuumQCD}
A widely used approach to developing a solution of QCD in the continuum is provided by the Dyson-Schwinger equations (DSEs) \cite{Roberts:1994dr, Chang:2011vu, Bashir:2012fs, Roberts:2015lja, Horn:2016rip, Eichmann:2016yit}, which define a symmetry-preserving [and hence Poincar\'e covariant] framework with a traceable connection to the Lagrangian of QCD.
The challenge in this approach is the need to employ a truncation in order to define a tractable bound-state problem.  
In this connection, much has been learnt in the past twenty years, so that one may now separate DSE predictions into three classes:
\emph{A}.\ model-independent statements about QCD;
\emph{B}.\ illustrations of such statements using well-constrained model elements and possessing a traceable connection to QCD;
\emph{C}.\ analyses that can fairly be described as QCD-based, but whose elements have not been computed using a truncation that preserves a systematically-improvable connection with QCD.

\begin{figure}[t]
\centerline{%
\includegraphics[clip,width=0.47\textwidth]{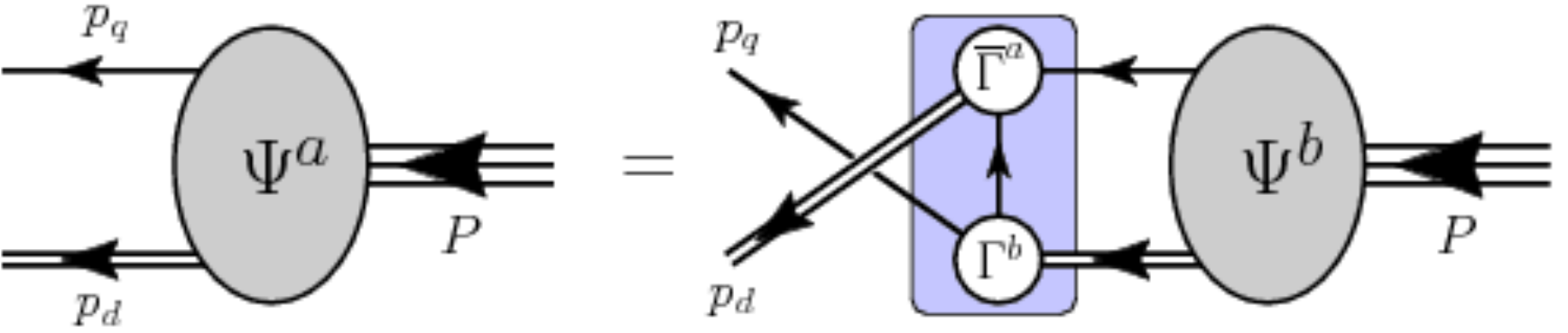}}
\caption{\label{figFaddeev}
Poincar\'e covariant Faddeev equation: a homogeneous linear integral equation for the matrix-valued function $\Psi$, being the Faddeev amplitude for a baryon of total momentum $P= p_q + p_d$, which expresses the relative momentum correlation between the dressed-quarks and -diquarks within the baryon.  The shaded rectangle demarcates the kernel of the Faddeev equation: \emph{single line}, dressed-quark propagator; $\Gamma$,  diquark correlation amplitude; and \emph{double line}, diquark propagator.  Further details are provided in Sec.\,\ref{sec:continuumQCD}.}
\end{figure}

The treatment of a baryon as a continuum three--valence-body bound-state problem became possible following the formulation of a Poincar\'e-covariant Faddeev equation \cite{Cahill:1988dx, Burden:1988dt, Cahill:1988zi, Reinhardt:1989rw, Efimov:1990uz}, which is depicted in Fig.\,\ref{figFaddeev}.  The ensuing years have seen studies increase in breadth and sophistication; and in order to understand those developments and the current status, it is apt to begin by elucidating the nature of the individual ``bodies'' whose interactions are described by that Faddeev equation.

It is worth opening with an observation, \emph{viz}.\ although it is commonly thought that the Higgs boson is the origin of mass, that is incorrect: it only gives mass to some very simple particles, accounting for just 1-2\% of the weight of more complex entities, such as atoms, molecules and everyday objects.  Instead, the vast bulk of all visible mass in the universe is generated dynamically by interactions in QCD \cite{Wilczek:2012sb}.  This remark is readily substantiated by noting that the mass-scale for the spectrum of strongly interacting matter is characterized by the proton's mass, $m_N \approx 1\,{\rm GeV} \approx 2000\,m_e$, where $m_e$ is the electron mass.  However, the only apparent scale in chromodynamics is the current-quark mass.  This is the quantity generated by the Higgs boson; but, empirically, the current-mass is just $1/250^{\rm th}$ of the scale for strong interactions, \emph{viz}.\ more-than two orders-of-magnitude smaller \cite{Olive:2016xmw}.  No amount of ``staring'' at the Lagrangian for QCD, ${\mathpzc L}_{\rm QCD}$, can reveal the source of that enormous amount of ``missing mass''.  Yet, it must be there;\footnote{This is a stark contrast to quantum electrodynamics [QED] wherein, \emph{e.g}.\ the scale in the spectrum of the hydrogen atom is set by $m_e$, which is a prominent feature of ${\mathpzc L}_{\rm QED}$ that is generated by the Higgs boson.}  and exposing the character of the Roper resonance is critical to understanding the nature of strong mass generation within the Standard Model.

One of the keys to resolving this conundrum is the phenomenon of DCSB \cite{Nambu:2011zz}, which can be exposed in QCD by solving the quark gap equation, \emph{i.e}.\ the Dyson-Schwinger equation [DSE] for the dressed-quark self-energy \cite{Roberts:1994dr}:
{\allowdisplaybreaks
\begin{subequations}
\label{gendseN}
\begin{align}
S^{-1}(p;\zeta) & = i\gamma\cdot p \, A(p^2;\zeta) + B(p^2;\zeta)\\
%
& = Z_2 \,(i\gamma\cdot p + m^{\rm bm}) + \Sigma(p;\zeta)\,,\\
\Sigma(p;\zeta)& =  Z_1 \int^\Lambda_{dq}\!\! g^2 D_{\mu\nu}(p-q;\zeta)\frac{\lambda^a}{2}\gamma_\mu S(q;\zeta) \Gamma^a_\nu(q,p;\zeta) ,
\end{align}
\end{subequations}}
\hspace*{-0.5\parindent}where the dressed-gluon propagator may be expressed via
\begin{equation}
\label{DressedGluon}
 D_{\mu\nu}(k;\zeta) = \Delta(k^2;\zeta) D^0_{\mu\nu}(k) \,,
\end{equation}
$k^2 D^0_{\mu\nu}(k^2)=\delta_{\mu\nu}- p_\mu p_\nu/p^2$;
$\Gamma_\nu^a=(\lambda^a/2) \Gamma_\nu$ is the gluon-quark vertex; $\int^\Lambda_{dq}$ indicates a Poincar\'e-invariant regularization of the integral, with regularization scale $\Lambda$; $m^{\rm bm}$ is the current-quark bare mass; and $Z_{1,2}(\zeta,\Lambda)$, respectively, are the vertex and quark wave-function renormalization constants, which also depend on the renormalization scale, $\zeta$.

The dressed-quark propagator in Eq.\,\eqref{gendseN} may be rewritten in the form:
\begin{equation}
\label{Mpdefinition}
S(p;\zeta) = Z(p^2;\zeta)/[i\gamma\cdot p + M(p^2)]\,,
\end{equation}
where $M(p^2) = B(p^2;\zeta) /A(p^2;\zeta) $ is the dressed-quark mass function, which is independent of $\zeta$.  In these terms, DCSB is the appearance of a $M(p^2) \not\equiv 0$ solution of Eq.\,\eqref{gendseN} when $m^{\rm bm}\equiv 0$, so that the quark acquires mass even in the absence of a Higgs mechanism.

Whether or not DCSB emerges in the Standard Model is decided by the structure of the gap equation's kernel.  Hence the basic question is: Just what form does that kernel take?  Owing to asymptotic freedom, the answer is known on the perturbative domain \cite{Jain:1993qh, Maris:1997tm, Maris:1999nt, Qin:2011dd, Qin:2011xq, Bloch:2002eq}, \emph{viz}.\  on ${\mathpzc A} = \{(p,q)\,|\, k^2=(p-q)^2 \simeq p^2 \simeq q^2 \gtrsim 2\,{\rm GeV}^2\}$:
\begin{equation}
\tfrac{g^2}{4\pi} D_{\mu\nu}(k) \, Z_1 \, \Gamma_\nu(q,p)
 \stackrel{{\mathpzc A}}{=} \alpha_s(k^2)\,
D^{0}_{\mu\nu}(k) \, Z_2^2 \,\gamma_\nu\,,
\label{UVmodelindependent}
\end{equation}
where $\alpha_s(k^2)$ is QCD's running coupling.   The question thus actually relates only to the infrared domain,  which is a complement of ${\mathpzc A}$, and so resides in sQCD.

The past two decades have revealed a great deal about the infrared behaviour of the running coupling, dressed-gluon propagator and dressed-gluon-quark vertex; and the current state of understanding can be traced from an array of sources \cite{Boucaud:2011ug, Binosi:2014aea, Aguilar:2015bud, Binosi:2016wcx, Binosi:2016nmeRMP}.  Of particular interest is the feature that the gluon propagator saturates at infrared momenta, \emph{i.e}.\
\begin{equation}
\Delta(k^2\simeq 0) = 1/m_g^2,
\end{equation}
which entails that the long-range propagation characteristics of gluons are dramatically affected by their self-interactions. Importantly, one may associate a renormalization-group-invariant (RGI) gluon mass-scale with this effect: $m_0 \approx 0.5\,$GeV\,$\approx m_N/2$ \cite{Binosi:2014aea, Binosi:2016nmeRMP, Cyrol:2016tym}, and summarize a large body of work, which began roughly thirty-five years ago \cite{Cornwall:1981zr}, by stating that gluons, although acting as massless degrees-of-freedom on the perturbative domain, actually possess a running mass, whose value at infrared momenta is characterised by $m_0$.

\begin{figure}[t]

\centerline{
\includegraphics[clip,width=0.42\textwidth]{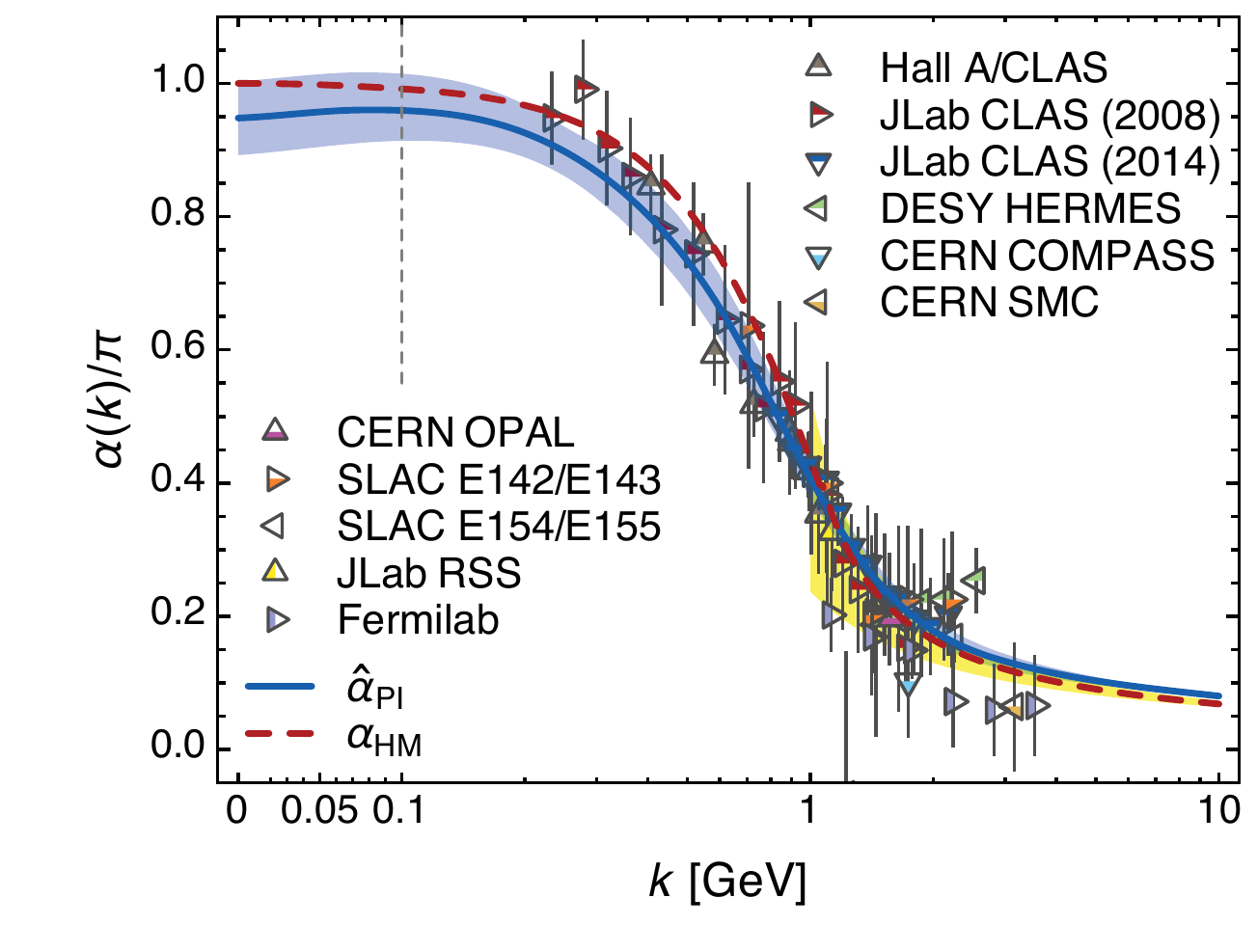}}

\caption{\label{FigwidehatalphaII}
Solid [blue] curve: process-in\-de\-pen\-dent RGI running-coupling $\widehat{\alpha}_{\rm PI}(k^2)$ \cite{Binosi:2016nmeRMP}.  The shaded (blue) band bracketing this curve combines a 95\% confidence-level window based on existing lQCD results for the gluon two-point function with an error of 10\% in the continuum analysis of relevant ghost-gluon dynamics.
World data on the process-dependent effective coupling $\alpha_{g_1}$, defined via the Bjorken sum rule \cite{%
Deur:2005cf, Deur:2008rf, Deur:2014vea,
Ackerstaff:1997ws, Ackerstaff:1998ja, Airapetian:1998wi, Airapetian:2002rw, Airapetian:2006vy,
Kim:1998kia,
Alexakhin:2006oza, Alekseev:2010hc, Adolph:2015saz,
Anthony:1993uf, Abe:1994cp, Abe:1995mt, Abe:1995dc, Abe:1995rn, Anthony:1996mw, Abe:1997cx, Abe:1997qk, Abe:1997dp, Abe:1998wq, Anthony:1999py, Anthony:1999rm, Anthony:2000fn, Anthony:2002hy}.
The shaded [yellow] band on $k>1\,$GeV represents $\alpha_{g_1}$ obtained from the Bjorken sum by using QCD evolution \cite{Gribov:1972, Altarelli:1977, Dokshitzer:1977} to extrapolate high-$k^2$ data into the depicted region \cite{Deur:2005cf, Deur:2008rf}; and, for additional context, the dashed [red] curve is the effective charge obtained in a light-front holographic model, canvassed elsewhere \cite{Deur:2016tte}.
%
}
\end{figure}

The mathematical tools that have enabled theory to arrive at this conclusion \cite{Abbott:1980hw, Abbott:1981ke, Cornwall:1981zr, Cornwall:1989gv, Pilaftsis:1996fh, Binosi:2002ft, Binosi:2003rr, Binosi:2009qm} can also be used to compute a \emph{process-independent} running-coupling for QCD, $\widehat{\alpha}_{\rm PI}(k^2)$  \cite{Binosi:2016nmeRMP}.  Depicted as the solid [blue] curve in Fig.\,\ref{FigwidehatalphaII}, this is a new type of effective charge, which is an analogue of the Gell-Mann--Low effective coupling in QED \cite{GellMann:1954fq} because it is completely determined by the gauge-boson propagator.  The result in Fig.\,\ref{FigwidehatalphaII} is a parameter-free Class-A prediction, capitalizing on analyses of QCD's gauge sector undertaken using both continuum methods and numerical simulations of lQCD.

The data in Fig.\,\ref{FigwidehatalphaII} represent empirical information on $\alpha_{g_1}$, a process-dependent effective-charge \cite{Grunberg:1982fw} determined from the Bjorken sum rule, one of the most basic constraints on our knowledge of nucleon spin structure.  Sound theoretical reasons underpin the almost precise agreement between $\widehat{\alpha}_{\rm PI}$ and $\alpha_{g_1}$ \cite{Binosi:2016nmeRMP}, so that the Bjorken sum may be seen as a near direct means by which to gain empirical insight into QCD's Gell-Mann--Low effective charge.
Given the behavior of the prediction in Fig.\,\ref{FigwidehatalphaII}, it is evident that the coupling is everywhere finite in QCD, \emph{i.e}. there is no Landau pole, and this theory possesses an infrared-stable fixed point.  Evidently, QCD is infrared finite owing to the dynamical generation of a gluon mass scale.\footnote{%
A theory is said to possess a Landau pole at $k^2_{\rm L}$ if the effective charge diverges at that point.  In QCD perturbation theory, such a Landau pole exists at $k^2_L=\Lambda_{\rm QCD}^2$.  Were such a pole to persist in a complete treatment of QCD, it would signal an infrared failure of the theory.  On the other hand, the absence of a Landau pole in QCD supports a view that QCD is unique amongst four-dimensional quantum field theories in being defined and internally consistent at all energy scales.  This might have implications for attempts to develop an understanding of physics beyond the Standard Model based upon non-Abelian gauge theories \cite{Appelquist:1996dq, Sannino:2009za, Appelquist:2009ka, Hayakawa:2010yn, Cheng:2013eu, Aoki:2013xza, DeGrand:2015zxa, Binosi:2016xxu}.}

As a unique process-independent effective charge, $\widehat{\alpha}_{\rm PI}$ appears in every one of QCD's dynamical equations of motion, setting the interaction strength in all cases, including the gap equation, Eq.\,\eqref{gendseN}.  It therefore plays a crucial role in determining the fate of chiral symmetry.

\begin{figure}[t]
\centerline{\includegraphics[width=0.38\textwidth]{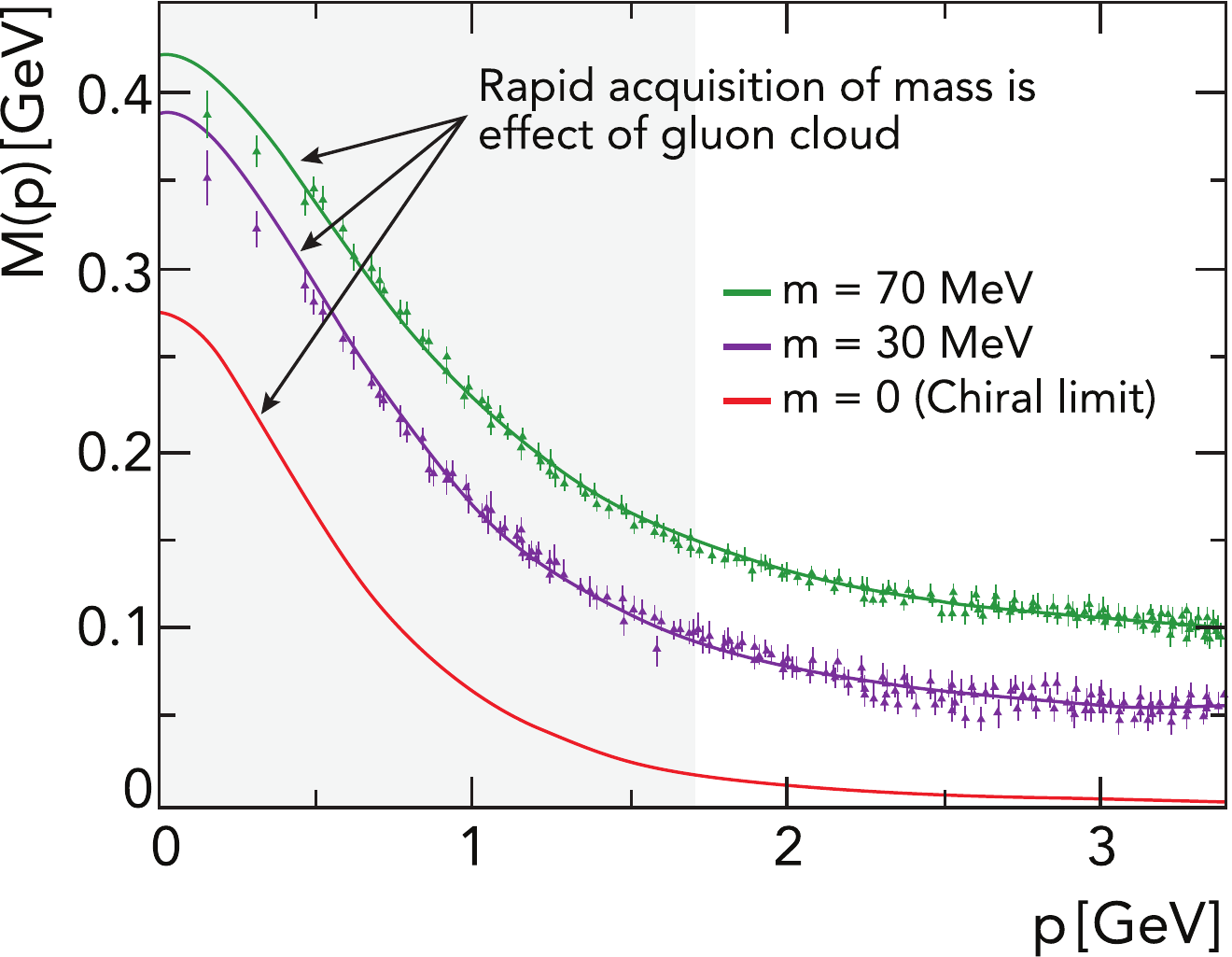}}
\caption{\label{gluoncloud}
Renormalization-group-invariant dressed-quark mass function, $M(p)$ in Eq.\,\eqref{Mpdefinition}: \emph{solid curves} -- gap equation results \cite{Bhagwat:2003vw,Bhagwat:2006tu}, ``data'' -- numerical simulations of lQCD \protect\cite{Bowman:2005vx}.  (\emph{N.B}.\ $m=70\,$MeV is the uppermost curve and current-quark mass decreases from top to bottom.)  The current-quark of perturbative QCD evolves into a constituent-quark as its momentum becomes smaller.  The constituent-quark mass arises from a cloud of low-momentum gluons attaching themselves to the current-quark.  This is DCSB, the essentially nonperturbative effect that generates a quark \emph{mass} \emph{from nothing}; namely, it occurs even in the chiral limit.
Notably, the size of $M(0)$ is a measure of the magnitude of the QCD scale anomaly in $n=1$-point Schwinger functions \cite{Roberts:2016vyn};
and experiments on $Q^2\in [0,12]\,$GeV$^2$ at the upgraded JLab facility will be sensitive to the momentum dependence of $M(p)$ within a domain that is here indicated approximately by the shaded region.
}
\end{figure}

The remaining element in the gap equation is the dressed gluon-quark vertex, $\Gamma_\nu$, whose complete expression involves twelve matrix-valued tensor structures, six of which are zero in the absence of chiral symmetry breaking.  If this vertex were only weakly modified from its tree-level form, $\gamma_\nu$, then, with $\widehat{\alpha}_{\rm PI}$ in Fig.\,\ref{FigwidehatalphaII}, chiral symmetry would be preserved in Nature \cite{Binosi:2016wcx}.  It is not; and after nearly forty years of studying $\Gamma_\nu$, with numerous contributions that may be traced from an analysis of Abelian theories \cite{Ball:1980ay}, continuum and lattice efforts have revealed just how the vertex is dressed so that DCSB is unavoidable.  Namely, the smooth, infrared-finite coupling depicted in Fig.\,\ref{FigwidehatalphaII} is strong enough to force nonzero values for the six terms in $\Gamma_\nu$ that usually vanish in the chiral limit.  This seeds a powerful positive feedback chain so that chiral symmetry is not only broken, but there is a sense in which it is very difficult to keep the growth of the dressed-quark mass function, $M(p^2)$, within physically reasonable bounds \cite{Binosi:2016wcx}.  Consequently, the solution of Eq.\,\eqref{gendseN} describes a dressed-quark with a dynamically generated running mass that is large in the infrared: $M(p^2 \simeq 0) \approx 0.3\,$GeV, as illustrated in Fig.\,\ref{gluoncloud}.

It is dressed quarks characterized by the mass function in Fig.\,\ref{gluoncloud} that are the basic elements in the Faddeev equation depicted in Fig.\,\ref{figFaddeev}, whose solutions in all allowed channels both generate the baryon spectrum and play a key role in computing the transitions between ground- and excited-states.
As highlighted elsewhere \cite{Cloet:2013gva, Binosi:2016wcx}, this means that since quarks carry electric charge, experiments involving electron scattering from hadrons serve as a probe of the momentum dependence of this mass function and also its collateral influences.  Measurements at the upgraded JLab facility will explore a region that is indicated approximately by the shading in Fig.\,\ref{gluoncloud}, \emph{i.e}.\ the domain of transition from strong- to perturbative-QCD.

Contemporary theory indicates that DCSB is responsible for more than 98\% of the visible mass in the Universe \cite{Brodsky:2015aiaRMP}.  Simultaneously, it ensures the existence of nearly-massless pseudo--Nambu-Goldstone modes [pions], each constituted from a valence-quark and -antiquark whose individual Lagrangian current-quark masses are $<1$\% of the proton mass \cite{Maris:1997hd}.

Another important consequence of DCSB is less well known.  Namely, any interaction capable of creating pseudo--Nambu-Goldstone modes as bound-states of a light dressed-quark and -antiquark, and reproducing the measured value of their leptonic decay constants, will necessarily also generate strong colour-antitriplet correlations between any two dressed quarks contained within a nucleon. 
Although a rigorous proof within QCD cannot be claimed, this assertion is based upon an accumulated body of evidence, gathered in two decades of studying two- and three-body bound-state problems in hadron physics \cite{Segovia:2015ufa}.  No realistic counter examples are known; and the existence of such diquark correlations is also supported by lQCD \cite{Alexandrou:2006cq, Babich:2007ah}.

The properties of such diquark correlations have been charted.  As color-carrying correlations, diquarks are confined \cite{Bender:1996bb, Bender:2002as, Bhagwat:2004hn}.  Additionally, owing to properties of charge-conjugation, a diquark with spin-parity $J^P$ may be viewed as a partner to the analogous $J^{-P}$ meson \cite{Cahill:1987qr}.  It follows that the strongest diquark correlations are: scalar isospin-zero, $[ud]_{0^+}$; and pseudovector, isospin-one, $\{uu\}_{1^+}$, $\{ud\}_{1^+}$, $\{dd\}_{1^+}$.  Moreover, whilst no pole-mass exists, the following mass-scales, which express the strength and range of the correlation, may be associated with these diquarks \cite{Cahill:1987qr, Maris:2002yu, Alexandrou:2006cq, Babich:2007ah, Eichmann:2016hgl, Lu:2017cln} [in GeV]:
\begin{equation}
m_{[ud]_{0^+}} \approx 0.7-0.8\,,\;
m_{\{uu\}_{1^+}}  \approx 0.9-1.1  \,,
\end{equation}
with $m_{\{dd\}_{1^+}}=m_{\{ud\}_{1^+}} = m_{\{uu\}_{1^+}}$ in the isospin symmetric limit.  The ground-state nucleon necessarily contains both scalar-isoscalar and pseudovector-isovector correlations: neither can be ignored and their presence has many observable consequences \cite{Roberts:2013mja, Segovia:2013uga}.

Realistic diquark correlations are also soft and interacting.  All carry charge, scatter electrons, and possess an electromagnetic size which is similar to that of the analogous mesonic system, \emph{e.g}.\ \cite{Maris:2004bp, Eichmann:2008ef, Roberts:2011wy}:
$r_{[ud]_{0^+}} \gtrsim r_\pi$, $r_{\{uu\}_{1^+}} \gtrsim r_\rho$,
with $r_{\{uu\}_{1^+}} > r_{[ud]_{0^+}}$.  As in the meson sector, these scales are set by that associated with DCSB.

It is important to emphasize that these fully dynamical diquark correlations are vastly different from the static, pointlike ``diquarks'' which featured in early attempts \cite{Lichtenberg:1967zz, Lichtenberg:1968zz} to understand the baryon spectrum and explain the so-called missing resonance problem, \emph{viz}.\ the fact that quark models predict many more baryons states than were observed in the previous millennium \cite{Burkert:2004sk}.   As we have stated, modern diquarks are soft [not pointlike].  They also enforce certain distinct interaction patterns for the singly- and doubly-represented valence-quarks within the proton, as reviewed elsewhere \cite{Roberts:2013mja, Segovia:2014aza, Roberts:2015lja, Segovia:2016zyc}.  On the other hand, the number of states in the spectrum of baryons obtained from the Faddeev equation \cite{Eichmann:2016hgl, Lu:2017cln} is similar to that found in the three-constituent quark model, just as it is in contemporary lQCD spectrum calculations \cite{Edwards:2011jj}.  [Notably, modern data and recent analyses have already reduced the number of missing resonances \cite{Ripani:2002ss, Burkert:2012ee, Kamano:2013iva, Crede:2013sze, Mokeev:2015moa,  Anisovich:2017pmi}.]

The existence of these tight correlations between two dressed quarks is the key to transforming the three valence-quark scattering problem into the simpler Faddeev equation problem illustrated in Fig.\,\ref{figFaddeev}, without loss of dynamical information \cite{Eichmann:2009qa}.   The three gluon vertex, a signature feature of QCD's non-Abelian character, is not explicitly part of the bound-state kernel in this picture.  Instead, one capitalizes on the fact that phase-space factors materially enhance two-body interactions over $n\geq 3$-body interactions and exploits the dominant role played by diquark correlations in the two-body subsystems.  Then, whilst an explicit three-body term might affect fine details of baryon structure, the dominant effect of non-Abelian multi-gluon vertices is expressed in the formation of diquark correlations.  Consequently, the active kernel here describes binding within the baryon through diquark breakup and reformation, which is mediated by exchange of a dressed-quark; and such a baryon is a compound system whose properties and interactions are largely determined by the quark$+$diquark structure evident in Fig.\,\ref{figFaddeev}.

This continuum approach to the baryon bound-state problem has been employed to calculate a wide range of nucleon-related observables \cite{Wilson:2011aa, Chang:2012cc, Roberts:2013mja, Segovia:2014aza, Roberts:2015dea, Xu:2015kta, Segovia:2016zyc, Eichmann:2016yit}.  In particular, in the computation of the mass and structure of the nucleon and its first radial excitation \cite{Segovia:2015hra}.  This Class-C analysis begins by solving the Faddeev equation, to obtain the masses and Poincar\'e-covariant wave functions for these systems,
taking each element of the equation to be as specified in \cite{Segovia:2014aza}, which provides a successful description of the properties of the nucleon and $\Delta$-baryon.  With those inputs, the masses are [in GeV]:
\begin{equation}
\label{eqMasses}
\mbox{nucleon\,(N)} = 1.18\,,\;
\mbox{nucleon-excited\,(R)} = 1.73\,.
\end{equation}

\begin{figure}[!t]
\centerline{%
\includegraphics[width=0.68\linewidth]{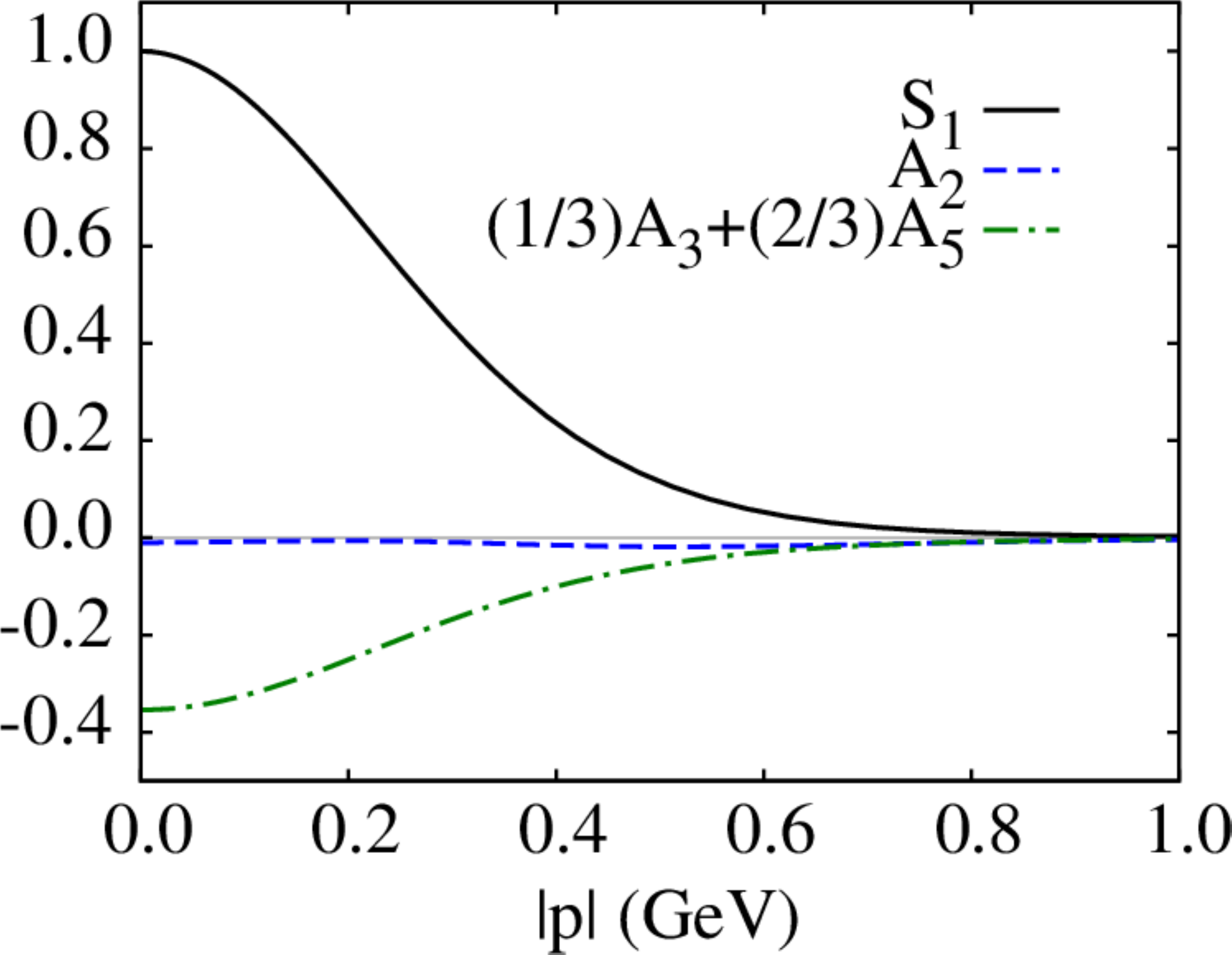}}
\vspace*{1ex}

\centerline{%
\includegraphics[width=0.68\linewidth]{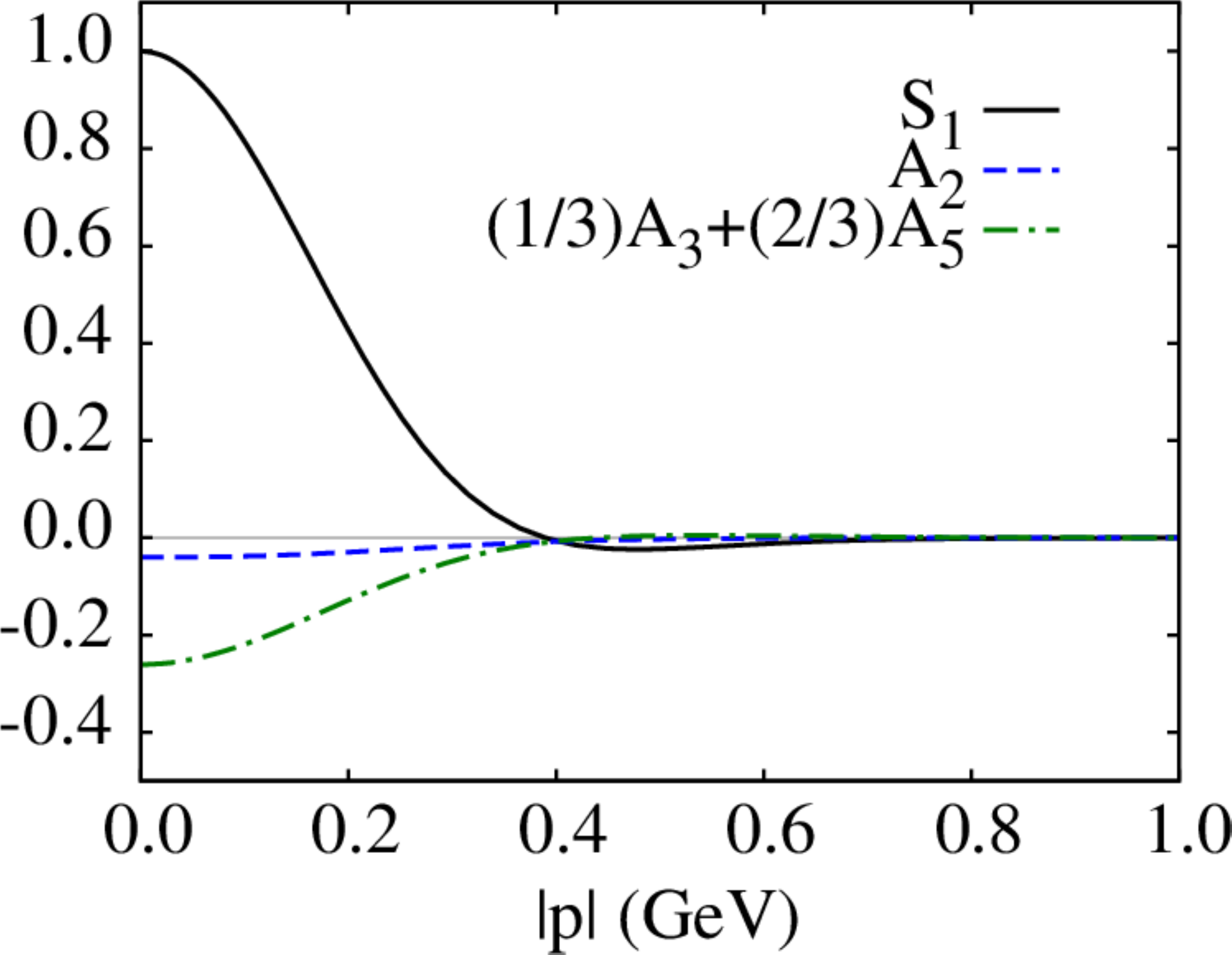}}
%
%
\caption{\label{figFA}
\emph{Upper panel}.  Zeroth Chebyshev moment of all $S$-wave components in the nucleon's Faddeev wave function, which is obtained from $\Psi$ in Fig.\,\ref{figFaddeev}, by reattaching the dressed-quark and -diquark legs.
\emph{Lower panel}.  Kindred functions for the first $J^P=1/2^+$ excited state.
Legend: $S_1$ is associated with the baryon's scalar diquark; the other two curves are associated with the axial-vector diquark; and here the normalization is chosen such that $S_1(0)=1$.}
\end{figure}

The masses in Eq.\,\eqref{eqMasses} correspond to the locations of the two lowest-magnitude $J^P=1/2^+$ poles in the three dressed-quark scattering problem.  The associated residues are the canonically-normalized Faddeev wave functions, which depend upon $(\ell^2,\ell \cdot P)$, where $\ell$ is the quark-diquark relative momentum and $P$ is the baryon's total momentum.  Figure\,\ref{figFA} depicts the zeroth Chebyshev moment of all $S$-wave components in that wave function, \emph{i.e}.\ projections of the form
\begin{equation}
{\mathpzc W}(\ell^2;P^2) = \frac{2}{\pi} \int_{-1}^1 \! du\,\sqrt{1-u^2}\,
{\mathpzc W}(\ell^2,u; P^2)\,,
\end{equation}
where $u=\ell\cdot P/\sqrt{\ell^2 P^2}$.  Drawing upon experience with quantum mechanics and with excited-state mesons studied via the Bethe-Salpeter equation \cite{Holl:2004fr, Qin:2011xq, Rojas:2014aka}, the appearance of a single zero in $S$-wave components of the Faddeev wave function associated with the first excited state in the three dressed-quark scattering problem indicates that this state is a radial excitation.  Notably, one may associate a four-vector length-scale of $1/[0.4 {\rm GeV}]\approx 0.5\,$fm with the location of this zero.
[Similar conclusions have been drawn using lQCD \cite{Roberts:2013ipa}.]

Let us focus now on the masses in Eq.\,\eqref{eqMasses}.  As discussed in connection with Fig.\,\ref{EBACRoper}, the empirical values of the pole locations for the first two states in the nucleon channel are: $0.939\,$GeV for the nucleon; and two poles for the Roper, $1.357 - i \,0.076$, $1.364 - i \, 0.105\,$GeV.  At first glance, these values appear unrelated to those in Eq.\,\eqref{eqMasses}.  However, deeper consideration reveals \cite{Eichmann:2008ae, Eichmann:2008ef} that the kernel in Fig.\,\ref{figFaddeev} omits all those resonant contributions which may be associated with the MB\,FSIs that are resummed in dynamical coupled channels models  \cite{JuliaDiaz:2007kz, Suzuki:2009nj, Kamano:2010ud, Ronchen:2012eg, Kamano:2013iva, Doring:2014qaa} in order to transform a bare-baryon into the observed state.  The  Faddeev equation analysed to produce the results in Eq.\,\eqref{eqMasses} should therefore be understood as producing the \emph{dressed-quark core} of the bound-state, not the completely-dressed and hence observable object.

Clothing the nucleon's dressed-quark core by including resonant contributions to the kernel produces a physical nucleon whose mass is $\approx 0.2$\,GeV lower than that of the core \cite{Ishii:1998tw, Hecht:2002ej, Chang:2009ae, Sanchis-Alepuz:2014wea}.  Similarly, clothing the $\Delta$-baryon's core lowers its mass by $\approx 0.16\,$GeV \cite{JuliaDiaz:2007kz}.   It is therefore no coincidence that [in GeV] $1.18-0.2 = 0.98\approx 0.94$, \emph{i.e}.\ the nucleon mass in Eq.\,\eqref{eqMasses} is 0.2\,GeV greater than the empirical value.  A successful body of work on the baryon spectrum \cite{Lu:2017cln}, and nucleon and $\Delta$ elastic and transition form factors \cite{Segovia:2014aza, Roberts:2015dea, Segovia:2016zyc} has been built upon this knowledge of the impact of omitting resonant contributions and the magnitude of their effects.  Therefore, a comparison between the empirical value of the Roper resonance pole-position and the computed dressed-quark core mass of the nucleon's radial excitation is not the critical test.  Instead, it is that between the masses of the quark core and the value determined for the meson-undressed bare-Roper, \emph{viz}.:
\begin{equation}
\label{eqMassesA}
\begin{array}{l|c}
    & \mbox{mass/GeV} \\\hline
\mbox{R}_{{\rm core}}^{\mbox{\footnotesize \cite{Segovia:2015hra}}} & 1.73\\
\mbox{R}_{{\rm core}}^{\mbox{\footnotesize \cite{Wilson:2011aa}}}   &  1.72\\
\mbox{R}_{{\rm core}}^{\mbox{\footnotesize \cite{Lu:2017cln}}}   &  1.82\\
\mbox{R}_{\rm DCC\,bare}^{\mbox{\footnotesize \cite{Suzuki:2009nj}}} & 1.76
\end{array}\,.
\end{equation}
Evidently, as already displayed in Fig.\,\ref{EBACRoper}, the DCC bare-Roper mass agrees with the quark core results obtained using both a QCD-kindred interaction \cite{Segovia:2015hra} and refined treatments of a strictly-implemented vector$\,\otimes\,$vector contact-interaction \cite{Wilson:2011aa, Lu:2017cln}.\footnote{It is also commensurate with the value obtained in simulations of lQCD whose formulation and/or parameters suppress MB\,FSIs, Fig.\,\ref{lQCDRoper1}.}  This is notable because all these calculations are independent, with just one common feature; namely, an appreciation that observed hadrons should realistically be built from a dressed-quark core plus a meson-cloud.

The agreement in Eq.\,\eqref{eqMassesA} is suggestive but not conclusive because, plainly, the same mass is obtained from the Faddeev equation using vastly different fundamental interactions.  The mass alone, then, does not serve as a fine discriminator between theoretical pictures of the nucleon's first radial excitation and its possible identification with the Roper resonance.   Critical additional tests are imposed by requiring that the theoretical picture combine a prediction of the Roper's mass with detailed descriptions of its structure and how that structure is revealed in the momentum dependence of the proton-Roper transition form factors.  Moreover, it must combine all this with a similarly complete picture of the proton, from which the Roper resonance is produced.  As detailed in Sec.\,\ref{Experiment}, precise empirical information is now available on the proton-Roper transition form factors, reaching to momentum transfers $Q^2\approx 4.5\,$GeV$^2$.  At such scales, these form factors probe a domain whereupon hard dressed-quark degrees-of-freedom could be expected to determine their behavior.  Finally, to increase the level of confidence, one should impose an additional test, requiring that the theoretical picture also explain all related properties of the $\Delta^+$-baryon, which is typically viewed as the proton's spin-flip excitation.

With Faddeev amplitudes for the participating states in hand, computation of the form factors in Eq.\,\eqref{NRcurrents} is a straightforward numerical exercise once the electromagnetic current is specified.  That sufficient to express the interaction of a photon with a baryon generated by the Faddeev equation in Fig.\,\ref{figFaddeev} is known \cite{Oettel:1999gc, Segovia:2014aza}.  It is a sum of six terms, with the photon separately probing the quarks and diquarks in various ways, so that diverse features of quark dressing and the quark-quark correlations all play a role in determining the form factors.

In any computation of transition form factors, one must first calculate the analogous elastic form factors for the states involved because the associated values of $F_1(Q^2=0)$ fix the normalization of the transition.  These normalizations can also be used to reveal the diquark content of the bound-states \cite{Roberts:2013mja, Segovia:2014aza, Segovia:2015hra}; and the analysis which produces the first row in Eq.\,\eqref{eqMassesA} yields:
\begin{equation}
\label{Pdiquark}
\begin{array}{l|cc}
        & N    & R    \\\hline
P_{J=0 \times 0} & 62\% & 62\% \\
P_{J\neq 0 \times 0} & 38\% & 38\% \\
\end{array}\,,
\end{equation}
where $P_{J=0 \times 0}$ measures the contribution to $F_1(Q^2=0)$ from overlaps with a scalar diquark in both the initial and final state, and $P_{J\neq 0 \times 0}$ is all the rest.  This calculation predicts that the relative strength of scalar and axial-vector diquark correlations in the nucleon and its radial excitation is the same.  However, the result is sensitive to the character of the quark-quark interaction.  Hence, this is a prediction that is tested by experiment.
Charge radii may also be computed from the elastic form factors, with the result \cite{Segovia:2015hra}: $r^{ \Psi}_{R^+}/r_{p}^{\Psi}=1.8$, \emph{i.e}.\ a quark-core radius for the radial excitation that is 80\% larger than that of the ground-state.  In contrast, non-relativistic harmonic oscillator wave functions yield a value of 1.5 for this ratio.  The difference highlights the impact of orbital angular momentum and spin-orbit repulsion, which is introduced by relativity into the Poincar\'e-covariant Faddeev wave functions for the nucleon and its radial excitation and increases the size of both systems.
The ratio of magnetic radii is $1.6$.

\begin{figure}[!t]
\includegraphics[height=0.52\linewidth,width=0.74\linewidth]{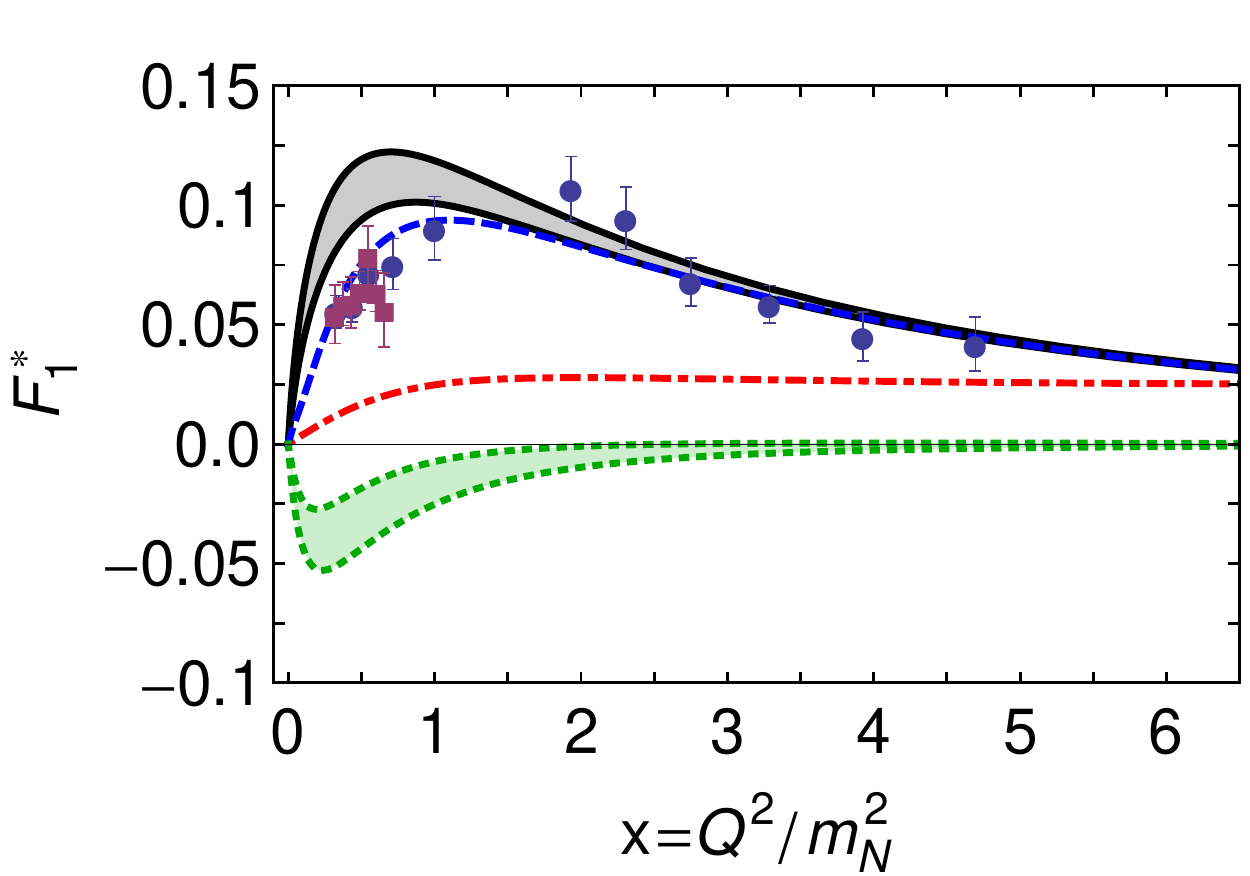}\hspace*{0.9em}
\vspace*{1ex}

\centerline{%
\includegraphics[width=0.72\linewidth]{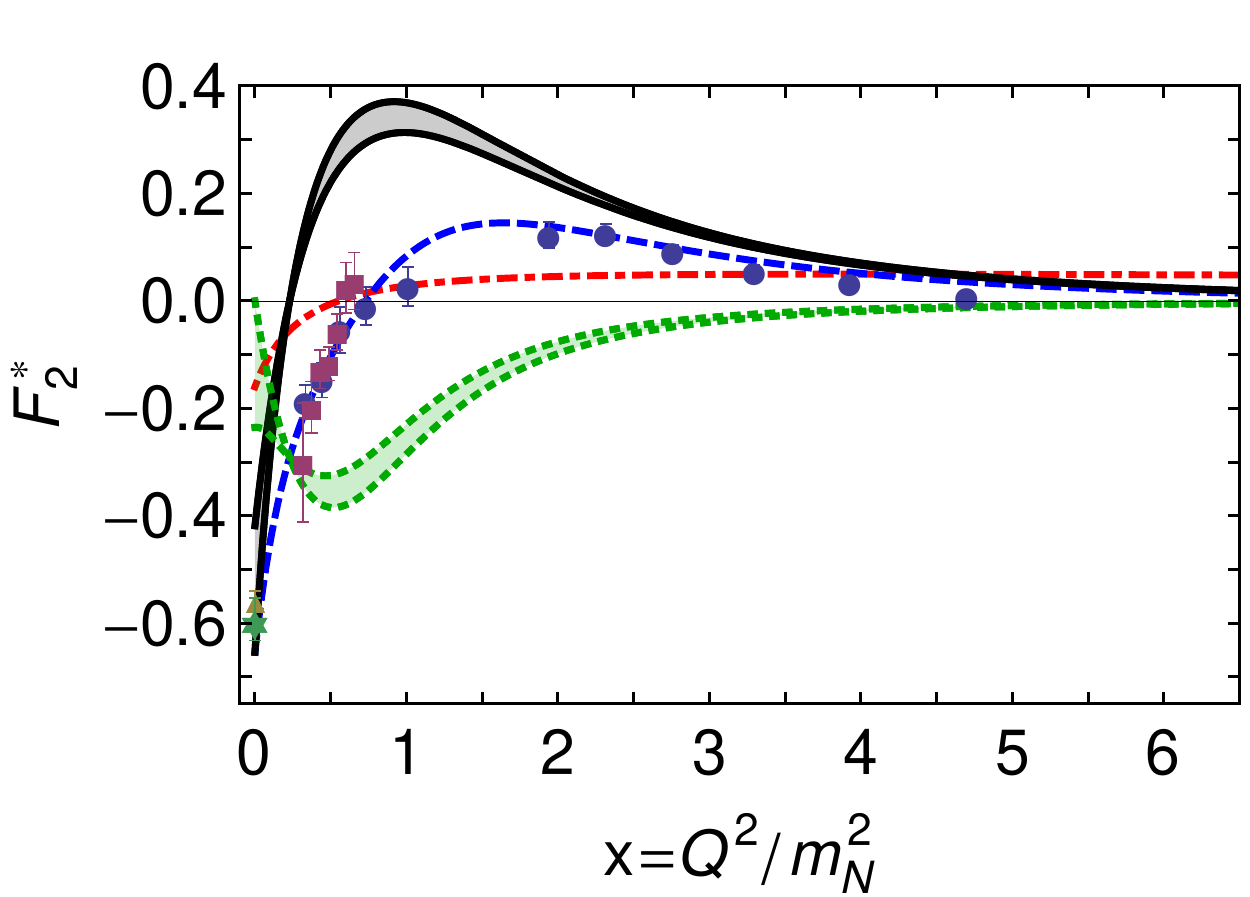}}
%
%
\caption{\label{figFT}
\emph{Upper panel} -- $F_{1}^{\ast}$ as a function of $x=Q^2/m_N^2$.
Legend: Gray band within black curves -- dressed-quark core contribution with up-to 20\% Faddeev amplitude renormalization from MB\,FSIs, implemented according to Eq.\,\eqref{eqMBFSI}.  The transition form factor curve with smallest magnitude at $x=1$ has the maximum renormalization.
Green band within green dotted curves -- inferred MB\,FSI contribution.  The band demarcates the range of uncertainty arising from $0\to 20$\% renormalization of the dressed-quark core.
Blue dashed curve --  least-squares fit to the data on $x \in (0,5)$.
Red dot-dashed curve -- contact interaction result \cite{Wilson:2011aa}.
\emph{Lower panel} -- $F_{2}^{\ast}(x)$ with same legend.
Data: circles [blue] \cite{Aznauryan:2009mx};
triangle [gold] \cite{Dugger:2009pn};
squares [purple] \cite{Mokeev:2012vsa, Mokeev:2015lda};
and star [green] \cite{Olive:2016xmw}.}
\end{figure}

The form factors predicted in \cite{Segovia:2015hra} to describe the transition between the proton and its first radial excitation are depicted in Fig.\,\ref{figFT}.
The upper panel depicts the Dirac transition form factor $F_{1}^{\ast}$, which vanishes at $x=Q^2/m_N^2 =0$ owing to orthogonality between the proton and its radial excitation.  The calculation [gray band] agrees quantitatively in magnitude and qualitatively in trend with the data on $x\gtrsim 2$.  Crucially, nothing was tuned to achieve these results.  Instead, the nature of the prediction owes fundamentally to the QCD-derived momentum-dependence of the propagators and vertices employed in formulating the bound-state and scattering problems.  This point is further highlighted by the contact-interaction result [red, dot-dashed]: with momentum-independent masses and vertices, the prediction disagrees both quantitatively and qualitatively with the data.  Experiment is evidently a sensitive tool with which to chart the nature of the quark-quark interaction and hence discriminate between competing theoretical hypotheses; and it is plainly settling upon an interaction that produces a momentum-dependent quark mass of the form in Fig.\,\ref{gluoncloud}, which characterises QCD.

The mismatch on $x\lesssim 2$ between data and the prediction in \cite{Segovia:2015hra} is also revealing.  As we have emphasized, that calculation yields only those form factor contributions generated by a rigorously-defined dressed-quark core whereas meson-cloud contributions are expected to be important on $x\lesssim 2$.  Thus, the difference between the prediction and data may plausibly be attributed to MB\,FSIs.  One can quantify this by recognizing that the dressed-quark core component of the baryon Faddeev amplitudes should be renormalized by inclusion of meson-baryon ``Fock-space'' components, with a maximum strength of 20\% \cite{Cloet:2008fw, Eichmann:2008ef, Bijker:2009up, Aznauryan:2016wwm}.  Naturally, since wave functions in quantum field theory evolve with resolving scale \cite{Lepage:1979zb, Lepage:1980fj, Efremov:1979qk, Raya:2015gva, Gao:2017mmp}, the magnitude of this effect is not fixed.  Instead ${\mathpzc I}_{MB}={\mathpzc I}_{MB}(Q^2)$, where $Q^2$ measures the resolving scale of any probe and ${\mathpzc I}_{MB}(Q^2) \to 0^+$ monotonically with increasing $Q^2$.  Now, form factors in QCD possess power-law behaviour, so it is appropriate to renormalize the dressed-quark core contributions via
\begin{subequations}
\begin{align}
\label{eqMBFSI}
F_{\rm core}(Q^2) & \to [1- {\mathpzc I}_{MB}(Q^2)] F_{\rm core}(Q^2)\,,\\
\quad {\mathpzc I}_{MB}(Q^2) & = [1-0.8^2]/[1+Q^2/\Lambda_{MB}^2]\,,
\end{align}
\end{subequations}
with $\Lambda_{MB}=1\,$GeV marking the midpoint of the transition between the strong and perturbative domains of QCD as measured by the behaviour of the dressed-quark mass-function in Fig.\,\ref{gluoncloud}.  Following this procedure \cite{Roberts:2016dnb}, one arrives at the estimate of MB\,FSI contributions depicted in Fig.\,\ref{figFT}.

The lower panel of Fig.\,\ref{figFT} depicts the Pauli form factor, $F_{2}^{\ast}$.  All observations made regarding $F_{1}^{\ast}$ also apply here, including those concerning the inferred meson-cloud contributions.  Importantly, the existence of a zero in $F_{2}^{\ast}$ is not influenced by meson-cloud effects, although its precise location is.

\begin{figure}[!t]
%
\includegraphics[height=0.52\linewidth,width=0.74\linewidth]{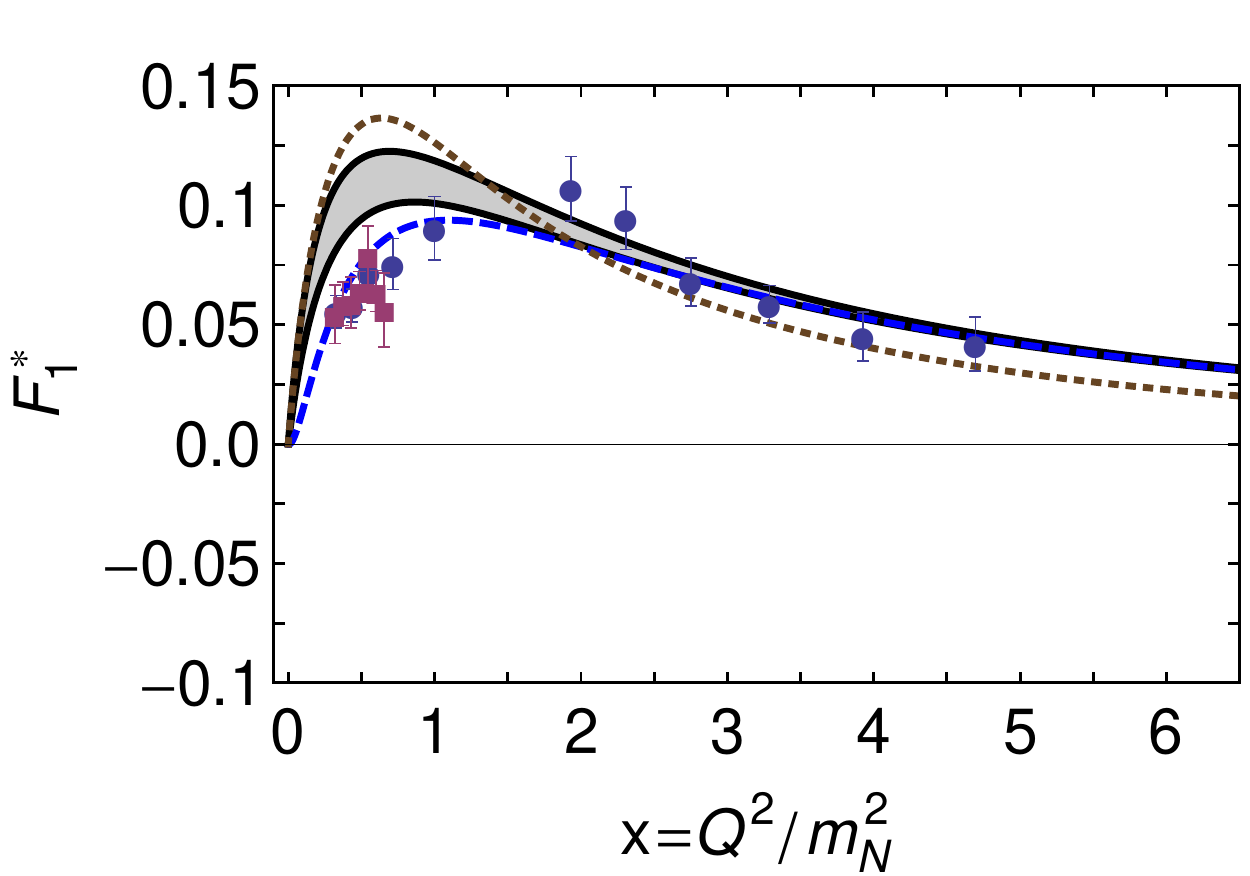}\hspace*{0.9em}
\vspace*{1ex}

\centerline{%
\includegraphics[width=0.72\linewidth]{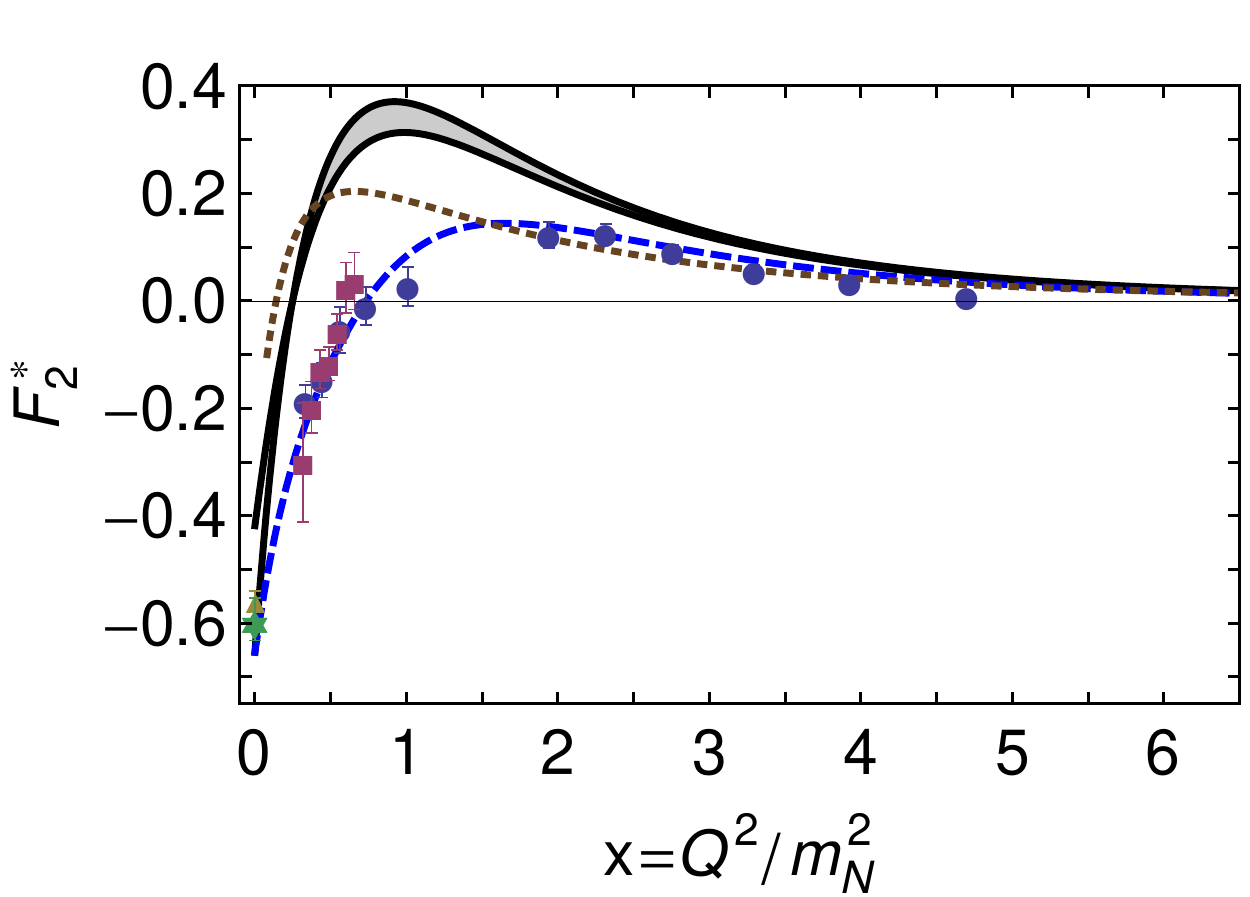}}
%
%
\caption{\label{aznauryanLFQM}
$F_{1}^{\ast}$ (upper panel) and $F_{2}^{\ast}$ (lower) for the proton-Roper transition as a function of $x=Q^2/m_N^2$.
Legend.  Gray band within black curves, prediction in \cite{Segovia:2015hra};
and dashed (blue) curve, least-squares fit to the data on $x \in (0,5)$.  [Both also depicted in Fig.\,\ref{figFT}.]
Dotted (brown) curve, LF\,CQM result reconstructed from the helicity amplitudes in \cite{Aznauryan:2016wwm} using Eqs.\,\eqref{ThHelAmp}.
Data: circles [blue] \cite{Aznauryan:2009mx};
triangle [gold] \cite{Dugger:2009pn};
squares [purple] \cite{Mokeev:2012vsa, Mokeev:2015lda};
and star [green] \cite{Olive:2016xmw}.
}
\end{figure}

This is an opportune moment to review the picture of the Roper resonance that is painted by constituent quark models.  Figure\,\ref{A12_lowQ} emphasized the importance of relativity in reproducing a zero in $F_2^\ast$, which generates the zero in $A_{1/2}$; and the discussion in this subsection has highlighted that the natural degrees-of-freedom to employ when studying measurable form factors are strongly-dressed quasi-particles (and correlations between them).  It is interesting, therefore, that constituent quark models, formulated using light-front quantization (LF\,CQMs) and incorporating aspects of the QCD dressing explained herein, have been used with success to describe features of the nucleon-Roper transition \cite{Cardarelli:1996vn, Aznauryan:2012ec, Aznauryan:2016wwm}.
In these models, the dressing effects are implemented phenomenologically, \emph{i.e}. via parametrizations chosen in order to secure a good fit to certain data; and they do not properly comply with QCD constraints at large momenta, \emph{e.g}.\ using constituent-quark electromagnetic form factors that fall too quickly with increasing momentum transfer \cite{Cardarelli:1996vn} or a dressed-quark mass function that falls too slowly \cite{Aznauryan:2012ec}.  Notwithstanding these limitations, the outcomes expressed are qualitatively significant.  This is illustrated in Fig.\,\ref{aznauryanLFQM}, which reveals a striking similarity between the DSE prediction for the dressed quark-core components of the transition form factors and those computed using a LF\,CQM that incorporates a running quark mass \cite{Aznauryan:2016wwm}.  The parameters of the LF\,CQM model were adjusted by fitting nucleon elastic form factors on $Q^2\in [0,16]\,$GeV$^2$, allowing room for MB\,FSIs and estimating their impact.   Qualitatively, therefore, despite fundamental differences in formulation, both the DSE and LF\,CQM approaches arrive at the same conclusion regarding the nature of the proton-Roper transition form factors: whilst MB\,FSIs contribute materially on $x\lesssim 2$, a dressed-quark core is exposed and probed on $x\gtrsim 2$.

It should be emphasized here that were the Roper a purely molecular meson-baryon system, in the sense defined in Sec.\,\ref{sec:DCC}, then the transition form factors would express an overlap between an initial state proton, which certainly possesses a dressed-quark core, and a much more diffuse system.  In such circumstances, $F_{1,2}^\ast$ would be far softer than anything that could be produced by a final state with a material dressed-quark core.  Consequently, the agreement between CLAS data and theory in Figs.\,\ref{figFT}, \ref{aznauryanLFQM} renders a molecular hypothesis untenable.

Finally, given the scope of agreement between experiment and theory in Figs.\,\ref{figFT}, \ref{aznauryanLFQM}, it is time to apply a final test, \emph{viz}.\ does the same perspective also deliver a consistent description of the nucleon and $\Delta$-baryon elastic form factors and the nucleon-$\Delta$ transition?  An affirmative answer is supported by an array of results \cite{Segovia:2014aza, Roberts:2015dea, Segovia:2016zyc}, from which we will highlight just one.

\begin{figure}[t]
\centerline{\includegraphics[clip,width=0.35\textwidth]{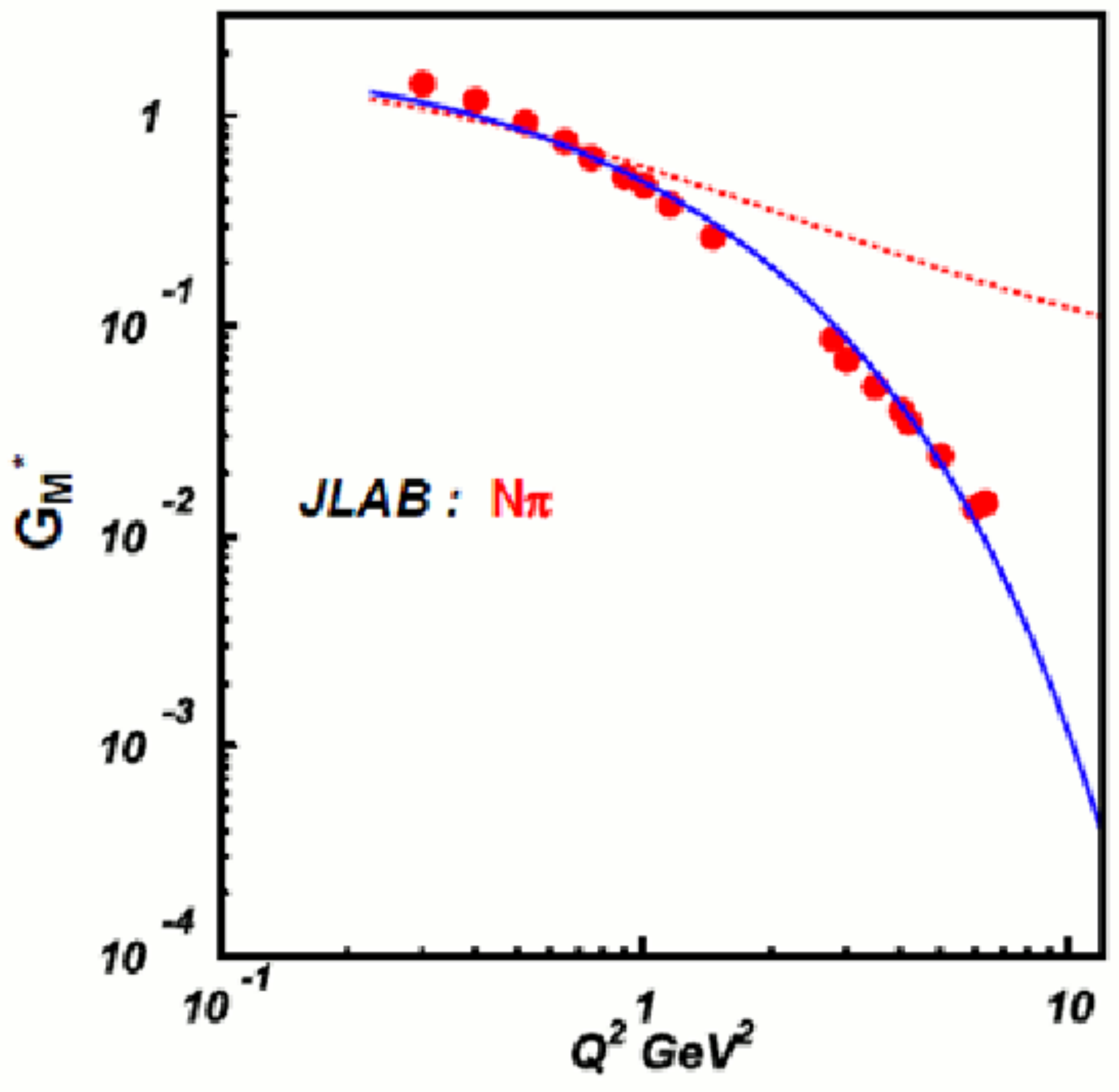}}
\caption{\label{figNDelta}
Comparison between data \cite{Aznauryan:2009mx} on the magnetic $\gamma^\ast+N\to\Delta$ transition form factor and a theoretical prediction (solid curve) \cite{Segovia:2014aza, Segovia:2016zyc}.  The dashed curve shows the result that would be obtained if the interaction between quarks were momentum-independent \cite{Segovia:2013rca}.}
\end{figure}

The $\gamma^\ast+N\to\Delta$ transition form factors excite keen interest because of their use in probing, \emph{inter} \emph{alia}, the relevance of perturbative QCD to processes involving moderate momentum transfers \cite{Carlson:1985mm, Pascalutsa:2006up, Aznauryan:2011qj}; shape deformation of hadrons \cite{Alexandrou:2012da}; and the role that resonance electroproduction experiments can play in exposing non-perturbative features of QCD \cite{Aznauryan:2012baS}.  Precise data on the dominant $\gamma^\ast+N\to\Delta$ magnetic transition form factor now reaches to $Q^2 = 6.5\,$GeV$^2$ \cite{Aznauryan:2009mx, Villano:2009sn}.  It poses both opportunities and challenges for QCD theory because this domain joins the infrared, where MB\,FSIs can be important, to the ultraviolet, where the dressed-quark core should control the transition.  The result obtained using the framework that produces the grey band in Figs.\,\ref{figFT}, \ref{aznauryanLFQM} is drawn as the solid curve in Fig.\,\ref{figNDelta}.  In this case, too, there is a mismatch between data and calculation at low-$Q^2$: both qualitatively and quantitatively, that difference can be attributed to MB\,FSIs \cite{Sato:2000jf, Burkert:2004sk, JuliaDiaz:2006xt, Pascalutsa:2006up, Crede:2013sze}.  However, on $Q^2\gtrsim 1\,$GeV$^2$ the theoretical curve agrees with the data: this is significant because, once again, no parameters were varied in order to ensure this outcome.  Importantly, a similar picture emerges when quark dressing effects are incorporated in LF\,CQMs  \cite{Aznauryan:2016wwm}.

\subsection{Light-Front Transverse Transition Charge Densities}
The nucleon-$\Delta$ and nucleon-Roper transition form factors have been dissected in order to reveal the relative contributions from dressed-quarks and the various diquark correlations \cite{Segovia:2016zyc}.  This analysis reveals that $F_1^\ast$ is largely determined by a process in which the virtual photon scatters from the uncorrelated $u$-quark with a $[ud]_{0^+}$ diquark as a spectator, with lesser but non-negligible contributions from other processes.  In exhibiting these properties, $F_1^\ast$ shows qualitative similarities to the proton's Dirac form factor.

Such features of the transition can also be highlighted by studying the following transition charge density: \cite{Tiator:2008kd}:
\begin{align}
\label{eqrhob}
\rho^{pR}(|\vec{b}|)
& := \int \frac{d^2 \vec{q}_\perp }{(2\pi)^2} \,{\rm e}^{i \vec{q}_\perp \cdot \vec{b}} F_1^\ast(|\vec{q}_\perp|^2)\,,
%
\end{align}
where $F_1^\ast$ is the proton-Roper Dirac transition form factor, depicted in Figs.\,\ref{figFT}, \ref{aznauryanLFQM} and interpreted in a frame defined by $Q=({q}_\perp=(q_1,q_2),Q_3=0,Q_4=0)$.  Plainly, $Q^2 = |\vec{q}_\perp|^2$.  Defined in this way, $\rho^{pR}(|\vec{b}|)$ is a light-front-transverse charge-density with a straightforward quantum mechanical interpretation \cite{Miller:2007uy}.  

\begin{figure}[t]
\centerline{%
\includegraphics[width=0.72\linewidth]{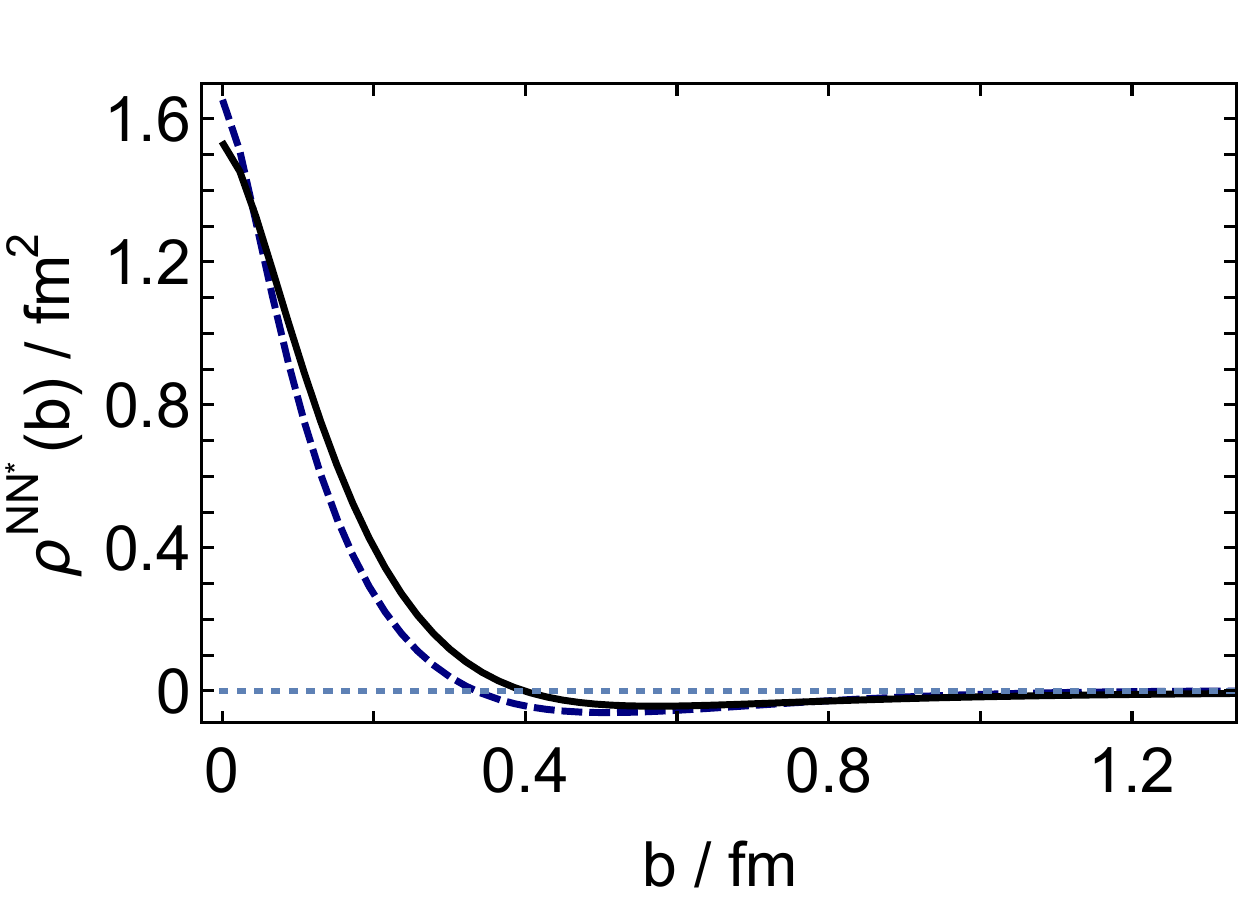}}
\centerline{%
\includegraphics[width=0.72\linewidth]{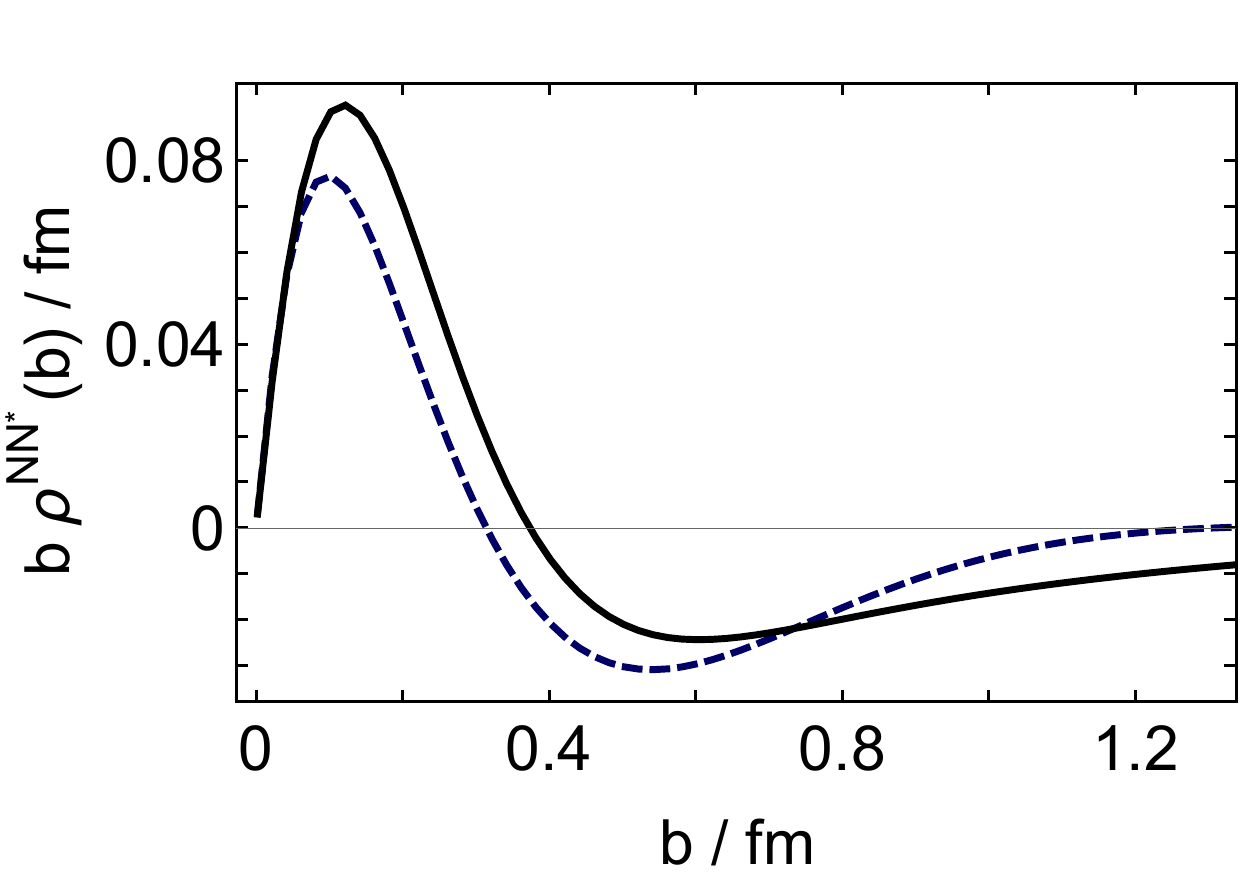}}
\caption{\label{rhob}
$\rho^{pR}(|\vec{b}|)$ (upper panel) and $|\vec{b}| \rho^{pR}(|\vec{b}|) $ (lower) calculated using Eq.\,\eqref{eqrhob}: solid (black) curve -- dressed-quark core contribution, computed using the midpoint-result within the gray bands in the left panel of Fig.\,\ref{figFT}; and dashed (blue) curve -- empirical result, computed using the dashed (blue) curve therein.}
\end{figure}

Fig.\,\ref{rhob} depicts a comparison between the empirical result for $\rho^{pR}(|\vec{b}|)$ and the dressed-quark core component: the difference between these curves measures the impact of MB\,FSIs on the transition.
Within the domain displayed, both curves describe a dense positive center, which is explained by noting that the proton-Roper transition is dominated by the photon scattering from a positively-charged $u$-quark in the presence of a positively-charged $[ud]_{0^+}$ diquark spectator, as mentioned above.
Furthermore, both curves exhibit a zero at approximately 0.3\,-\,0.4\,fm, with that of the core lying at larger $|\vec{b}|$.  Thence, after each reaching a global minimum, the dressed-quark core contribution returns slowly to zero from below whereas the empirical result returns to pass through zero once more, although continuing to diminish in magnitude.

The long-range negative tail of the dressed-quark core contribution, evident in Fig.\,\ref{rhob}, reveals the increasing relevance of axial-vector diquark correlations at long range because the $d\{uu\}_{1^+}$ component is twice as strong as $u\{ud\}_{1^+}$ in the proton and Roper wave functions, and photon interactions with uncorrelated quarks dominate the transition.
Moreover, consistent with their role in reducing the nucleon and Roper quark-core masses, one sees that MB\,FSIs introduce significant attraction, working to screen the long negative tail of the quark-core contribution and thereby compressing the transition domain in the transverse space.  [The dominant long-range MB effect is $n\pi^+$, which generates a positive tail.] In fact, as measured by the rms transverse radius, the size of the empirical transition domain is just two-thirds of that associated with the dressed-quark core.

\section{Conclusion}
After more than fifty years, a coherent picture connecting the Roper resonance with the nucleon's first radial excitation has become visible.  Completing this portrait only became possible following
the acquisition and analysis of a vast amount of high-precision nucleon-resonance electroproduction data with single- and double-pion final states on a large kinematic domain of energy and momentum-transfer,
development of a sophisticated dynamical reaction theory capable of simultaneously describing all partial waves extracted from available, reliable data,
formulation and wide-ranging application of a Poincar\'e covariant approach to the continuum bound state problem in relativistic quantum field theory that expresses diverse local and global impacts of DCSB in QCD,
and the refinement of constituent quark models so that they, too, qualitatively incorporate these aspects of strong QCD.
In this picture:
\begin{itemize}
\setlength\itemsep{0em}
\item the Roper resonance is, at heart, the first radial excitation of the nucleon.
\item It consists of a well-defined dressed-quark core, which plays a role in determining the system's properties at all length-scales, but exerts a dominant influence on probes with $Q^2\gtrsim m_N^2$, where $m_N$ is the nucleon mass;
\item and this core is augmented by a meson cloud, which both reduces the Roper's core mass by approximately 20\%, thereby solving the mass problem that was such a puzzle in constituent-quark model treatments, and, at low-$Q^2$, contributes an amount to the electroproduction transition form factors that is comparable in magnitude with that of the dressed-quark core, but vanishes rapidly as $Q^2$ is increased beyond $m_N^2$.
\end{itemize}

These fifty years of experience with the Roper resonance have delivered lessons that cannot be emphasized too strongly. Namely, in attempting to predict and explain the QCD spectrum, one must: fully consider the impact of meson-baryon final-state interactions (MB\,FSIs), and the couplings between channels and states that they generate; and look beyond merely locating the poles in the $S$-matrix, which themselves reveal little structural information, to also consider the $Q^2$-dependences of the residues, which serve as a penetrating scale-dependent probe of resonance composition.

Moreover, the Roper resonance is not unusual.  Indeed, in essence, the picture drawn here is also applicable to the $\Delta$-baryon; and an accumulating body of experiment and theory indicates that almost all baryon resonances can be viewed the same way, \emph{viz}.\ as systems possessing a three-body dressed-quark bound-state core that is supplemented by a meson cloud, whose importance varies from state to state and whose observable manifestations disappear rapidly as the resolving power of the probe is increased.  In this connection, it is important to highlight that CLAS12 at the newly upgraded JLab will be capable of determining the electrocouplings of most prominent nucleon resonances at unprecedented photon virtualities: $Q^2\in [6,12]\,$GeV$^2$ \cite{E12-09-003, E12-06-108A}.  Consequently, the associated experimental program will be a powerful means of validating the perspective described herein.

Assuming the picture we've drawn is correct, then CLAS12 will deliver empirical information that can address a wide range of issues that are critical to our understanding of strong interactions, \emph{e.g}.: is there an environment sensitivity of DCSB; and are quark-quark correlations an essential element in the structure of all baryons?  As reviewed herein, existing experiment-theory feedback suggests that there is no environment sensitivity for the nucleon, $\Delta$-baryon and Roper resonance: DCSB in these systems is expressed in ways that can readily be predicted once its manifestation is understood in the pion, and this includes the generation of diquark correlations with the same character in each of these baryons.
Resonances in other channels, however, may contain additional diquark correlations, with different quantum numbers, and potentially be influenced in new ways by MB\,FSIs.  Therefore, these channels, and higher excitations, open new windows on sQCD and its emergent phenomena whose vistas must be explored and mapped if the most difficult part of the Standard Model is finally to be solved.

\section*{Acknowledgments}
In preparing this article we benefited greatly from constructive comments and input provided by
I.\,G.~Aznauryan,
A.~Bashir,
D.~Binosi,
L.~Chang,
C.~Chen,
Z.-F.~Cui,
B.~El-Bennich,
R.~Gothe,
L.\,X.~Guti\'errez-Guerrero,
G.~Krein,
T.-S.\,H.~Lee,
H.-W.~Lin,
K.-F.~Liu,
Y.~Lu,
C.~Mezrag,
V.~Mokeev,
J.~Papavassiliou,
J.-L.~Ping,
H.\,L.\,L.~Roberts,
J.~Rodrig\'{\i}uez-Quintero,
T.~Sato,
S.\,M.~Schmidt,
J.~Segovia,
S.-S.~Xu,
F.~Wang,
D.\,J.~Wilson,
and H.-S.~Zong.
This work was supported by
U.S.\ Department of Energy, Office of Science, Office of Nuclear Physics, under contract nos.~DE-AC05-06OR23177 and DE-AC02-06CH11357.





\end{document}